\documentclass[twocolumn,aps,prl,superscriptaddress,floatfix]{revtex4-2}
\usepackage{graphicx,subfigure}
\usepackage{amsmath,amssymb,bm}
\usepackage{mathtools}
\usepackage{multirow}
\usepackage{makecell}
\usepackage{textcomp}
\usepackage{float}
\usepackage{color}
\usepackage[normalem]{ulem}
\usepackage{dsfont}
\usepackage{mathrsfs}
\usepackage{braket}
\usepackage{transparent}
\usepackage{xcolor}
\usepackage{hhline}
\usepackage{graphicx}
\usepackage{pdfpages}
\usepackage[export]{adjustbox}
\usepackage{lineno}
\usepackage{hyperref}
\usepackage{physics}
\usepackage{comment}

\makeatletter
\AtBeginDocument{\let\LS@rot\@undefined}
\makeatother

\newcommand{\figOne}{
 \begin{figure*}[t]
    \centering
    \includegraphics[width=\textwidth]{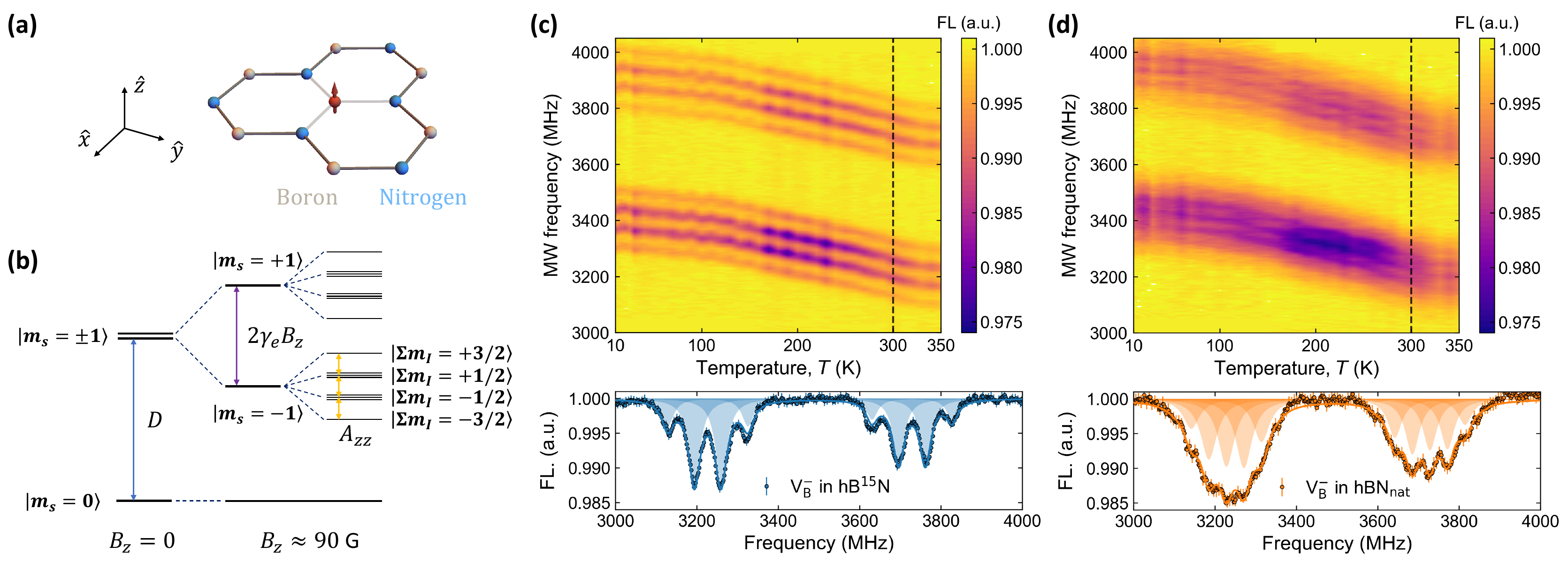}
    \caption{{\bf Temperature dependence of optically detected magnetic resonance spectra of \vbm.}
    (a) Schematic representation of a single \vbm center (red spin) in the hBN honeycomb lattice. $\hat{z}$ is defined as the out-of-plane direction, while $\hat{x}$ and $\hat{y}$ are in the lattice plane.
    (b) The \vbm electronic ground state energy level diagram with the presence of three nearest \nfive nuclear spins in isotopically purified \hbn flakes. 
    The $|m_s=\pm1\rangle$ is separated from $|m_s=0\rangle$ by a zero-field splitting $D$.
    The hyperfine interaction further splits each spin transition into four transitions with spacing $A_{zz}$ and degeneracy of \{1, 3, 3, 1\}.
    (c) ODMR spectrum of \vbm in \hbn flakes under magnetic field $B_z \approx 90$~G at temperatures ranging from 10~K to 350~K.
    The normalized fluorescence (FL) contrast is marked in the colobar.
    The bottom panels of (c) and (d) display the ODMR spectrum at 300~K, corresponding to the black dashed line on the top panels.
    The spectrum is fitted with two groups of equally spaced Lorentzians to extract the values of ZFS and hyperfine splitting.
    (d) ODMR spectrum of \vbm in \hbnnat flakes in the 10-350~K temperature range.
    }
    \label{fig:fig1}
 \end{figure*}
}

\newcommand{\figTwo}{
 \begin{figure}[t]
    \centering
    \includegraphics[width=\linewidth]{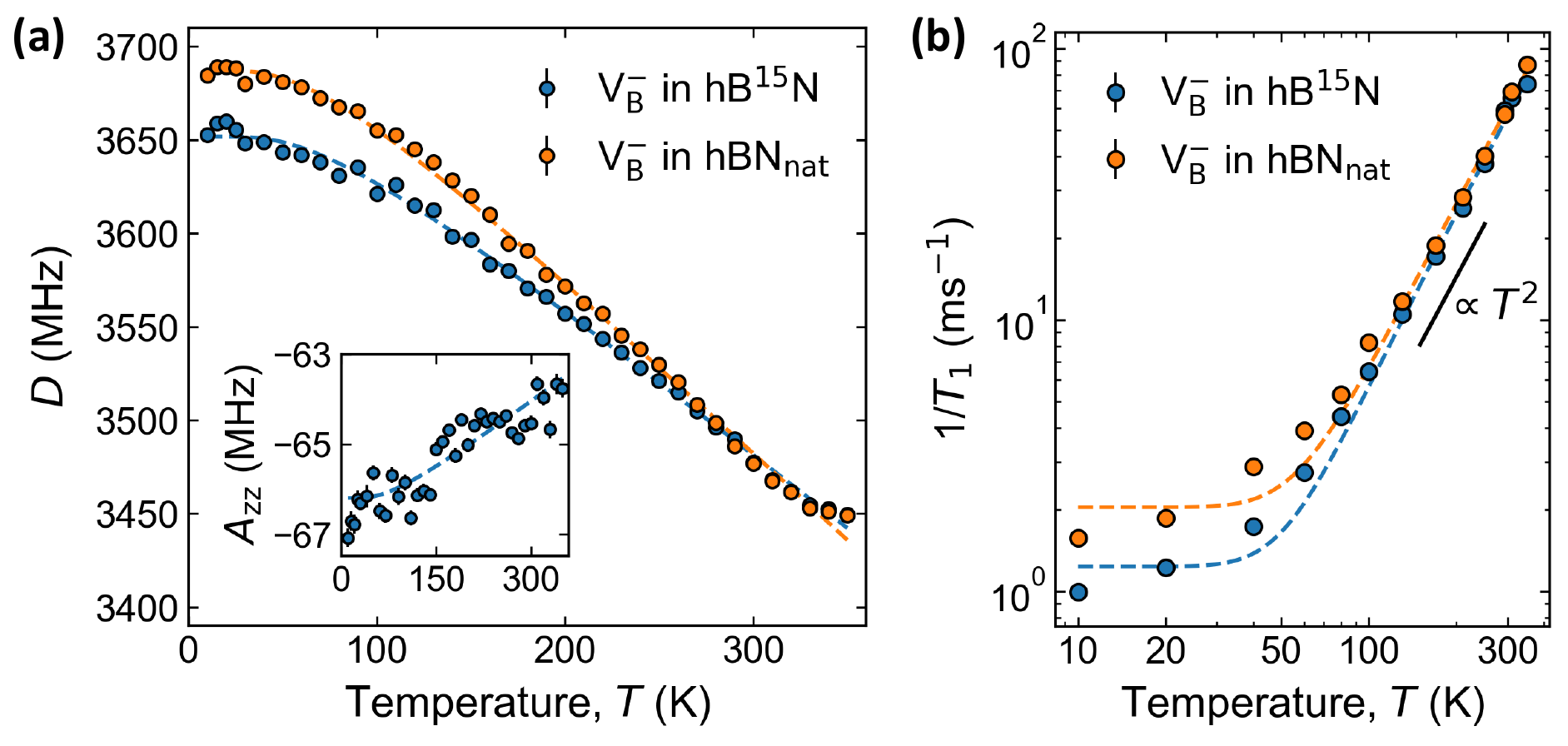}
    \caption{{\bf Temperature-dependent properties of \vbm in different hBN samples.}
    (a) The temperature dependence of the ZFS $D(T)$ of \vbm in the range 10-350~K.
    The dotted lines represent a fit to a physically motivated model using Eq.~\ref{eq:physModel}.
    The errorbar at each points represents $1\sigma$ fitting error of the ODMR spectrum.
    We note that the errorbars are hardly visible due to the large change of the ZFS.
    Inset: hyperfine interaction $A_{zz}(T)$ of \vbm in \hbn within the same temperature range, fitted by the same model with one fixed phonon energy $18.4$~meV extracted from the fit of the ZFS.    
    (b) Spin relaxation rate $1/T_1$ of \vbm in the temperature range 10-350~K.
    The errorbar represents $1\sigma$ fitting error of the $T_1$ decay, which is also hardly visible due to the dramatic change.
    Setting the phonon energy at $\hbar\omega_\mathrm{exp}$, the dotted lines qualitatively reproduce the $T_1$ temperature-dependence by the model Eq.~\ref{eq:t1}.
    In the high temperature regime, the relaxation rate is approximated by a power-law scaling $1/T_1 \propto T^2$.
    }
    \label{fig:fig2}
 \end{figure}
}

\newcommand{\figThree}{
 \begin{figure}[h]
    \centering
    \includegraphics[width=\linewidth]{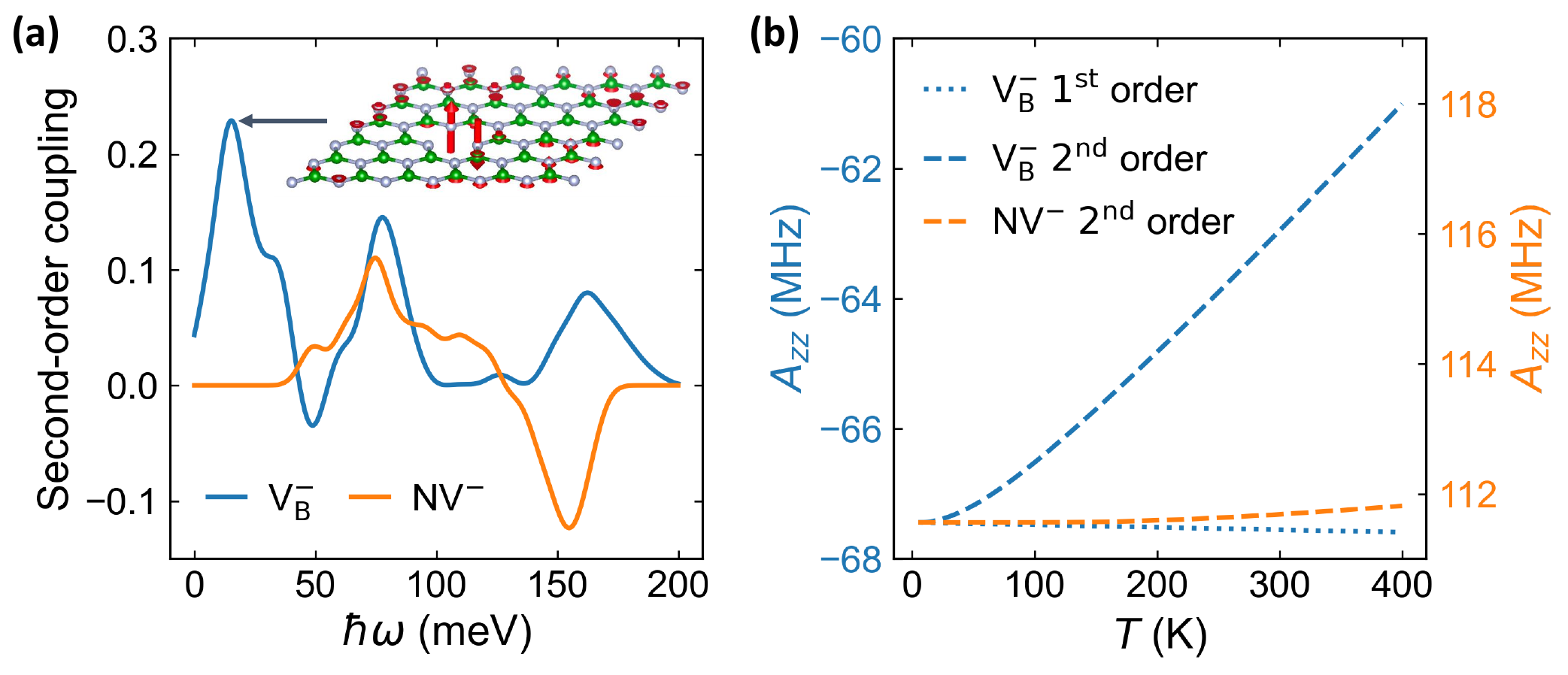}
    \caption{{\bf First principle simulations.}
    (a) The second-order vibrational coupling, i.e., $\frac{\partial^2 A_{\mathrm{zz}}}{\partial q_i^2} \frac{\hbar}{M_i \omega_i}$ as a function of phonon frequencies of the nearest \nfive in hBN (blue) and $^{13}$C in diamond (orange) to the vacancy, respectively. The first peak of \vbm in hBN is identified as an out-of-plane vibrational mode (inset).
    (b) The computed $A_\mathrm{zz}(T)$ of the nearest \nfive in hBN (blue) from both the first and second order vibrational contribution, respectively. We also plotted the computed $A_\mathrm{zz}(T)$ of the nearest $^{13}$C in NV$^-$ in diamond (orange) from the second-order vibrational contribution as a comparison.
    }
    \label{fig:fig3}
 \end{figure}
}

\newcommand{\figFour}{
 \begin{figure}[t]
    \centering
    \includegraphics[width=\linewidth]{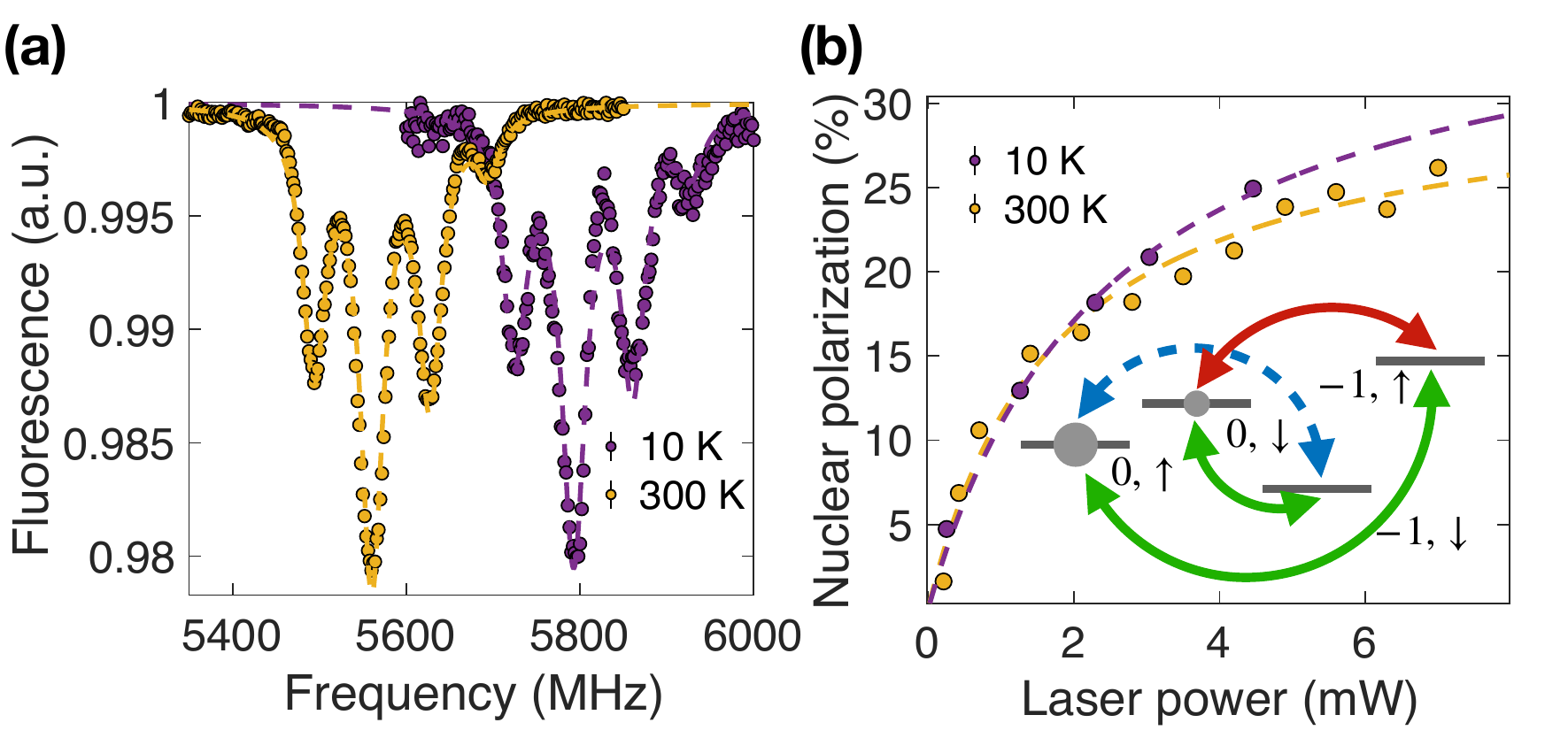}
    \caption{{\bf Dynamic nuclear polarization at low temperature.}
    (a) Near esLAC level ODMR Spectra of the $|m_s = 0\rangle \leftrightarrow |m_s = -1\rangle$ transition at 10 K and 300 K. The spectra both exhibit similar asymmetry toward the left peaks, indicating a polarization of nuclear spins (see Supplemental Material)
    (b) Polarization level of the three nearest \nfive nuclear spins near esLAC level as a function of the normalized laser power under room/cryogenic temperature. Inset: schematic representation of the polarization process. The strong spin-conserving optical polarization (green arrow) continuously pumps state from $|m_s=-1\rangle$ to $|m_s=0\rangle$, and the two hybridisation processes (red arrow and dashed blue arrow) differ in strength, resulting in the polarization of $|m_I = \:\uparrow\rangle$ state (see Supplemental Material).
    }
    \label{fig:fig4}
 \end{figure}
}

\newcommand{\tableOne}{
 \begin{table}[b]
    \caption{ {\bf Zero-field splitting fitting parameters.}
    For each isotope, the extracted parameters are averaged over four sets of data in different sample spacial positions (see Supplemental Material), fitted with same model in Eq.~\ref{eq:physModel}.
    We note that $D_0$ differs from $D(T=0)$, where $D_0 = D(T=0)-\frac{1}{2}c_D$.
    The \nfive and \nfour isotope effect of zero-field splitting $D$ has also been reported in NV centers~\cite{lourette2023temperature}.
    }
    \begin{ruledtabular}
        \begin{tabular}{ c c c c }
            \textrm{Isotopes}&
            \textrm{$D_0$ (MHz)}&
            \textrm{$c_D$ (MHz)}&
            \textrm{$\hbar\omega_\mathrm{exp}$ (meV)}\\ [0.2ex]
            \colrule
            \\ [-2ex]
            \nfive & $3742\pm10$ & $-175\pm13$ & $18.4\pm1.0$\\
            \nfour & $3777\pm26$ & $-201\pm42$ & $18.8\pm2.8$\\
        \end{tabular}
    \end{ruledtabular}
    \label{tab:table1}
 \end{table}
}

\usepackage{graphicx}

\graphicspath{ {figures/} }

\makeatletter
\let\saved@includegraphics\includegraphics
\AtBeginDocument{\let\includegraphics\saved@includegraphics}
\makeatother

\makeatletter
\newcommand*{\centerfloat}{%
  \parindent \z@
  \leftskip \z@ \@plus 1fil \@minus \textwidth
  \rightskip\leftskip
  \parfillskip \z@skip}
\makeatother

\newcommand{\vbm}[0]{$\mathrm{V}_{\mathrm{B}}^-$ }
\newcommand{\vbmns}[0]{$\mathrm{V}_{\mathrm{B}}^-$}

\newcommand{\hbn}[0]{$\mathrm{h}{}^{10}\mathrm{B}{}^{15}\mathrm{N}$ }
\newcommand{\hbnnat}[0]{$\mathrm{hBN_\mathrm{nat}}$ }
\newcommand{\hbnns}[0]{$\mathrm{h}{}^{10}\mathrm{B}{}^{15}\mathrm{N}$}
\newcommand{\hbnnatns}[0]{$\mathrm{hBN_\mathrm{nat}}$}

\newcommand{\nfour}[0]{${}^{14}\mathrm{N}$ }
\newcommand{\nfive}[0]{${}^{15}\mathrm{N}$ }

\begin{document}

\title{Temperature Dependent Spin-Phonon Coupling of Boron-Vacancy Centers in Hexagonal Boron Nitride}

\author{
Zhongyuan~Liu,$^{1,*}$ 
Ruotian~Gong,$^{1,*}$ 
Benchen~Huang,$^{2,*}$ 
Yu~Jin,$^{2}$ 
Xinyi~Du,$^{1}$
Guanghui~He,$^{1}$\\
Eli~Janzen,$^{3}$
Li~Yang,$^{1}$
Erik Henriksen,$^{1}$
James Edgar,$^{3}$
Giulia Galli,$^{2,4,5}$
Chong~Zu$^{1,6,7,\dag}$
\\
\medskip
\normalsize{$^{1}$Department of Physics, Washington University, St.~Louis, MO 63130, USA}\\
\normalsize{$^{2}$Department of Chemistry, University of Chicago, IL 60637, USA}\\
\normalsize{$^{3}$Tim Taylor Department of Chemical Engineering, Kansas State University, Manhattan, KS, 66506, USA}\\
\normalsize{$^{4}$Pritzker School of Molecular Engineering, University of Chicago, IL 60637, USA}\\
\normalsize{$^{5}$Material Science Division, Argonne National Lab, IL 60439, USA}\\
\normalsize{$^{6}$Center for Quantum Leaps, Washington University, St. Louis, MO 63130, USA}\\
\normalsize{$^{7}$Institute of Materials Science and Engineering, Washington University, St. Louis, MO 63130, USA}\\
\normalsize{$^*$These authors contributed equally to this work}\\
\normalsize{$^\dag$To whom correspondence should be addressed; E-mail: zu@wustl.edu}\\
}

\begin{abstract}
The negatively charged boron-vacancy center (\vbmns) in hexagonal boron nitride (hBN) has recently emerged as a highly promising quantum sensor. 
Compared to the nitrogen-vacancy (NV) center in diamond, the change with temperature of the spin transition energy of \vbm is more than an order of magnitude larger, making it a potential nanoscale thermometer with superior sensitivity.
However, the underlying mechanism of the observed large temperature dependence remains an open question.
In this work, using isotopically purified \hbnns, we systematically characterize the zero-field splitting, hyperfine interaction, and spin relaxation time of \vbm from 10 to 350~K.
We carry out first-principle calculations of the \vbm spin-phonon interaction and show that a second-order effect from finite-temperature phonon excitations is responsible for the observed changes in experiments.
By fitting our experimental results to a physically motivated model, we extract the dominant phonon mode which agrees well with our simulations.
Finally, we investigate the dynamic nuclear spin polarization process at cryogenic temperatures.
Our results provide key insights in \vbm centers and their utilization as nanoscale thermometers and phonon sensors. 
\end{abstract}

\date{\today}

\maketitle

\emph{Introduction}--- 
Optically-addressable solid state spin defects are promising platforms for quantum applications \cite{kolkowitz2012coherent,doherty2013nitrogen,aharonovich2016solid, awschalom2018quantum, atature2018material, wolfowicz2021quantum, togan2010quantum,pompili2021realization,degen2017quantum,zu2021emergent,koehl2011room,nagy2019high, hensen2015loophole, randall2021many, hsieh2019imaging, thiel2019probing, davis2023probing, mittiga2018imaging,he2023quasi, dwyer2022probing,Bhattacharyya2024-ko,acosta2009diamonds,de2021materials,luo2024room}, and when residing in atomically-thin van der Waals materials they may exhibit properties superior to those of their conterparts in three-dimensional materials~\cite{azzam2021prospects, ren2019review, caldwell2019photonics,naclerio2023review, ye2019spin}.
Among a wide-range of contestants, the negatively charged boron-vacancy center, \vbmns, in hexagonal boron nitride (hBN) has attracted much growing interest \cite{gottscholl2020initialization,gottscholl2021room,gong2023coherent, huang2022wide,broadway2020imaging, kumar2022magnetic, healey2023quantum, robertson2023detection, gao2023quantum,udvarhelyi2023planar,hennessey2023framework,patrickson2024high, ivady2020ab,kianinia2020generation}.
Several interesting properties of the host material, including the large bandgap and excellent stability, as well as the readily controllable spin degree of freedom at room temperature, makes \vbm in hBN well-suited as a platform for quantum technologies.

When compared with its three-dimensional counterparts, for example the nitrogen-vacancy (NV) center in diamond, the ground state zero-field splitting (ZFS) of \vbm exhibits a $\sim25$ times larger variation from cryogenic to room temperatures, placing it in an advantageous position for nanoscale thermometry \cite{vaidya2023quantum, liu2021temperature, gottscholl2021spin, kucsko2013nanometre, neumann2013high}.
However, the underlying mechanism of such a large temperature response remains an open question.
In particular, the first-order lattice displacement of hBN with temperature can only account for a negligible fraction of the changes in the measured ZFS of \vbm~\cite{gottscholl2021spin}; higher-order spin-phonon interactions should be included to quantitatively capture the experimental results.

In this letter, we experimentally characterize the temperature dependence of the ZFS, hyperfine interaction, and spin relaxation time $T_1$ of \vbm in isotopically purified \hbnns. 
Compared to conventional hBN with natural abundant $^{14}$N, our \hbn offers much narrower \vbm spin transitions and substantially improved measurement resolution~\cite{gong2024isotope, janzen2023boron, clua2023isotopic, sasaki2023nitrogen}.
We then carry out first-principle calculations of the temperature-induced phonon-mediated interaction of \vbm,  motivated by recent investigations of the NV centers in diamond~\cite{cambria2023physically, tang2023first}.
Our combined experimental and theoretical study allows for the identification of the main mechanism responsible for the observed temperature dependence of the properties of \vbmns, namely a second-order spin-phonon coupling with a characteristic phonon mode energy around $18~$meV.
We also investigate the dynamic nuclear spin polarization of the nearest three $^{15}$N and find that the polarization persists down to $10~$K.

\figOne

\emph{Experimental system}--- 
The \vbm is an atomic spin defect in the hBN lattice (Fig.~\ref{fig:fig1}a).
The electronic ground state of \vbm exhibits a spin-1 triplet with $|m_s = 0\rangle$ and $|m_s = \pm1\rangle$.
In the absence of external fields, the $|m_s = \pm1\rangle$ are degenerate and separated from $|m_s = 0\rangle$ by a zero-field splitting (ZFS) $D = (2\pi)\times 3.48~$GHz at room temperature (Fig.~\ref{fig:fig1}b).

The Hamiltonian governing \vbm and the nearest three nitrogen nuclear spins takes the form, 
\begin{equation}
\label{eq:H_total}
\begin{split}
    \mathcal{H} 
    &= D S_z^2
     + \gamma_e B_z S_z
     + \sum_{j=1}^{3}\mathbf{S}\mathbf{A}^j\mathbf{I}^j
     - \sum_{j=1}^{3}\gamma_n B_z I_{z}^j    ,
\end{split}
\end{equation}
where $B_z$ is an external magnetic field along the c-axis of hBN (or $\hat{z}$ direction), $\mathbf{S}$ and $S_z$ are the electron spin-1 operators, $\mathbf{I}^j$ and $I_z^j$ are the nuclear spin operators, and
$\mathbf{A}^j$ is the hyperfine tensor coupling electronic and nuclear spins.
The factors $\gamma_e= (2\pi)\times 2.8~$MHz/G and $\gamma_n$ are electronic and nuclear gyromagnetic ratios.

Throughout our experiment, the external magnetic field is $B_z \lesssim 760~$G.
In this case, the energy splittings between $|m_s = 0\rangle$ and $|m_s = \pm1\rangle$ sub-levels, $D \pm \gamma_e B_z$, are much larger than the strength of the hyperfine interaction $|\mathbf{A}^j|$, thus the secular approximation can be adopted: 
$\sum_{j}\mathbf{S}\mathbf{A}^j\mathbf{I}^j \approx \sum_{j}A_{zz}^j S_z I_z^j = A_{zz}(\sum_{j}I_z^j)S_z$, leading to a nuclear spin dependent energy shift of the electronic transition.
This results in $2\mathcal{I}+1$ transitions between $|m_s=0\rangle$ and $|m_s=-1\rangle$ (or $|m_s=+1\rangle$) sub-levels, where $\mathcal{I} = \sum_j I^j$ is the total nuclear spin number accounting for all three nearest nitrogen atoms (Fig.~1b).

The corresponding spin transitions of \vbm can be probed via optically detected magnetic resonance (ODMR) spectroscopy: by sweeping the frequency of the applied microwave drive while monitoring the fluorescence signal of \vbmns, we expect a fluorescence drop whenever the microwave is resonant with an electronic spin transition.
We report measurements for hBN with naturally distributed isotopes \hbnnat (99.6\% $^{14}\mathrm{N}$ with $I = 1$), and for the isotope-engineered \hbn (99.7\% $^{15}\mathrm{N}$ with $I=\frac{1}{2}$)~\cite{gong2024isotope,janzen2023boron}.
For conventional \hbnnatns, there are in total seven broad hyperfine spin transitions; while for \hbnns, we observe four sharp resonances, enabling the measurement of ZFS and hyperfine splitting, $A_{zz}$, with substantially improved resolution (Fig.~1cd).

\emph{Experiment results}---
To investigate the temperature-dependent spin properties of \vbmns, we load hBN samples with thickness ranging from $\sim 30 - 80$ nm into a closed-cycle optical cryostat (Fournine Design) for temperature control from $10-350~$K.
A small external magnetic field $B_z \approx 90$~G is used to lift the degeneracy between $|m_s=\pm1\rangle$ to suppress the local electric field effect~\cite{gong2023coherent,udvarhelyi2023planar} and better resolve the hyperfine interaction.
In principle, one can also perform thermometry with zero external magnetic field.
Figure~\ref{fig:fig1}c,d show the ODMR spectra as a function of temperature.
We fit the spectra with two groups of equally spaced Lorentzians, corresponding to $|m_s=0\rangle\leftrightarrow|m_s=-1\rangle$ and $|m_s=0\rangle\leftrightarrow|m_s=+1\rangle$ transitions respectively.
The ZFS, $D(T)$, can be extracted from the average frequency of the two transition groups, while the hyperfine splitting, $A_{zz}(T)$, is obtained from the fitted spacing between adjacent resonances in each group (Fig.~\ref{fig:fig2}a).
In addition, we also perform spin relaxation measurements and extract the $T_1$ timescale of the sample under different temperatures  (Fig.~\ref{fig:fig2}b).

\figTwo

We find that the ZFS of \vbm exhibits a dramatic change of more than $200$~MHz from 10 to 350 K, which is around 30 times larger than the change observed for NV centers~\cite{cambria2023physically}.
We determine the susceptibility of \vbm at room temperature, $\chi(T=300~\mathrm{K})$, to be $-0.89\pm0.04$~MHz/K for \hbnnatns, and $-0.77\pm0.03$~MHz/K for \hbn.
By using the susceptibility, we estimate the temperature sensitivity of \vbm in \hbn $\eta_T \approx 0.37~ \mathrm{K}/\sqrt{\mathrm{Hz}}$ at room temperature, outperforming the NV centers in nanodiamonds with $\sim100~$nm size (see Supplemental Material)~\cite{fujiwara2020real, tzeng2015time,choi2020probing,fujiwara2021diamond, gu2023simultaneous,bradac2020optical}.
These results highlight the potential use of \vbm as an ultra-sensitive nanoscale thermometer.
Furthermore, within our measured temperature range, we observe a slightly smaller change of ZFS in \hbn compared to \hbnnatns, which can be ascribed to the fact that the heavier nuclei in \hbn lead to atomic displacements less sensitive to the temperature change.
The isotope effect is further evidenced by the measured $T_1$ timescales, limited by spin-phonon interaction~\cite{gottscholl2021room, durand2023optically, cambria2023temperature}: for heavier nuclei and thus weaker spin-phonon coupling strength, the $T_1$ of \vbm in \hbn is longer at lower temperatures.

We also note that while it remains a challenge to resolve the hyperfine interaction strength from the broad ODMR spectra of conventional \hbnnat~\cite{gong2024isotope}, the substantially better resolved resonances of \hbn enable an accurate characterization of $A_{zz}(T)$ (see Supplemental Material). 
The measured amplitude of $A_{zz}(T)$ displays a noticeable increase from $64~$MHz to $67~$MHz with decreasing temperature, whereas the hyperfine interaction of NV centers to the nearby nuclear spins has been measured to be nearly constant with temperature~\cite{barson2019temperature, xu2023high}.
Finally, we observe that the ODMR contrast of \vbm in both hBN samples reaches a maximum around $210~$K and persists to low temperatures, without any substantial quenching. 
This behaviour is again opposite to that observed for the NV center in diamond, whose contrast suffers from a significant drop below $90~$K~\cite{fischer2013optical, ernst2023temperature, happacher2023temperature}, hindering its sensing performance at cryogenic temperatures.

\tableOne

\emph{Theoretical model}--- To understand the observed temperature dependence of \vbm spin properties, we adopt a theoretical model which was originally developed to study NV centers in diamond~\cite{tang2023first}.
In particular, we write the electronic and nuclear spin transitions as the sum of two terms,
\begin{equation}
    \nu = \nu_0(a(T)) + \frac{1}{2} \sum_i \frac{\partial^2 \nu}{\partial q_i^2} \frac{\hbar}{M_i \omega_i} \left(\frac{1}{e^{\hbar \omega_i/k_B T} - 1} + \frac{1}{2}\right),
\end{equation}
where $\nu$ is the transition frequency due to, e.g., ZFS and hyperfine interaction, and $a(T)$ is the lattice constant of hBN as a function of temperature~\cite{gottscholl2021spin}, $q_i, M_i, \omega_i$ are the normal mode of the crystal, mode-specific effective mass, and frequency, respectively.
Here the first-order term, $\nu_0(a(T))$, corresponds to the thermal expansion of the hBN lattice, while the second-order term represents vibrational contributions caused by finite-temperature phonon excitations near the equilibrium geometry.

We carry out first-principle simulations using density functional theory (DFT)~\cite{kohn1965self} with the Perdew-Burke-Ernzerhof (PBE) functional~\cite{perdew1996generalized} to compute the temperature-dependent hyperfine interaction, $A_{zz}(T)$, between \vbm and the nearest three \nfive nuclei. 
Interestingly, we find that the variation of the lattice constant with temperature only results in a negligible change in the magnitude of $A_{zz}(T)$.
On the other hand, the second-order term accounts for the effect of temperature on the value of $A_{zz}(T)$ observed experimentally.

Figure~\ref{fig:fig3}a shows the calculated second-order derivative of the energy of the crystal for each phonon mode.
We find a phonon mode with the largest amplitude at energy $\hbar\omega_\mathrm{th}\approx16$ meV, corresponding to an out-of-plane vibration shown in the inset.
The computed temperature dependence of $A_{zz}(T)$ agrees well with our experimental results (Fig.~\ref{fig:fig3}b).
For a direct comparison with the NV center, we also investigate the phonon mode associated with the hyperfine coupling between the NV center and its first shell $^{13}$C nuclear spin~\cite{tang2023first}.
In the case of the NV center, the dominant phonon-mode energy is $\simeq$ $70~$meV~\cite{cambria2023temperature, liu2023first}, requiring a much higher activation temperature than \vbm.
Hence the variation of $A_{zz}$ with temperature is much larger for the \vbm center  than that observed  for the NV center within the range ($10-350$~K) investigated in our experiment.

The dominant phonon mode found theoretically at $\hbar\omega_\mathrm{th}\approx16~$meV can also be independently determined by fitting the experimentally measured spin properties to a physically motivated model~\cite{cambria2023physically},
\begin{equation}
    \nu(T) = \nu_{0} + c_\nu\left(\frac{1}{e^{\hbar\omega/k T} - 1} + \frac{1}{2}\right),
    \label{eq:physModel}
\end{equation}
where $\nu(T=0) = \nu_0+\frac{1}{2}c_\nu$ is the transition energy at 0~K and $n = (e^{\hbar\omega/k T} - 1)^{-1}$ is the occupation number of the phonon mode.

We apply this model to the measured ZFS and hyperfine interaction of \vbmns.
For ZFS, the observed temperature dependence of $D(T)$ for both \hbn and \hbnnat samples can be accurately fitted using the model described in Eq.~\ref{eq:physModel}, from which we extract the characteristic phonon energy to be $\hbar\omega_\mathrm{exp} = 18.4\pm1.0~$meV and $18.8\pm2.8~$meV respectively (Fig.~\ref{fig:fig2}a and Table~\ref{tab:table1}), in agreement with our theoretical results.
The measured hyperfine interaction, $A_{zz}(T)$, shows a temperature dependence that is also consistent with the model when the phonon energy is set to $18.4$~meV (Fig.~\ref{fig:fig2}a Inset).
Interestingly, the agreement between $\hbar\omega_\mathrm{exp}$ and the calculated $\hbar\omega_\mathrm{th}$ suggests that the temperature dependence of ZFS and of the hyperfine interaction originate from the coupling of the electronic spin with vibrational mode identified in our calculations.

We now turn to investigate the temperature dependence of the spin relaxation time $T_1$.
Within the temperature range ($10-350$~K) considered here, $T_1$ of \vbm is dominated by Raman scattering processes, where the energy difference between phonon absorption/emission is equal to the ZFS of \vbmns, leading to spin depolarization (see Supplemental Material)~\cite{cambria2023temperature}.
Therefore the spin relaxation rate of \vbm can be approximated as~\cite{cambria2023temperature}
\begin{equation}
    \Gamma(T) = 1/T_1 \approx A n (n + 1) + A_S,
    \label{eq:t1}
\end{equation}
where $n = (e^{\hbar\omega/k T} - 1)^{-1}$ and $A$ are the phonon occupation number and coupling coefficient associated with the effective mode, and $A_S$ is a sample-related constant that might vary slightly between different experiments.
Using the phonon energy $\hbar\omega_\mathrm{exp} = 18.4\; (18.8)~$meV determined above, the model of Eq.~\ref{eq:t1} faithfully reproduces the experimentally measured spin relaxation rates from $60-350~$K (Fig.~\ref{fig:fig2}b).
The discrepancy below $60~$K may be due to the activation of lower-energy phonon modes as well as processes beyond Raman scattering. 
We note that a prior theoretical study has predicted a power-law scaling of $T_1\sim T^{-2}$~\cite{mondal2023spin}, which is consistent with our experimental results in the high temperature regime (Fig.~\ref{fig:fig2}b).

\figThree

\emph{Dynamic nuclear polarization at low temperature}---
With the temperature-dependent \vbm spin properties in hand, we now turn to investigate the dynamical polarization of the three nearest-neighbor \nfive nuclear spins at low temperature.
Nuclear spins feature exceptional isolation from noisy environments, making them ideal candidates for quantum storage applications~\cite{bradley2019ten, metsch2019initialization, bourassa2020entanglement}.
At room temperature, several prior works have demonstrated the polarization of proximate $^{14}$N or $^{15}$N with an upper limit of $\sim 30~\%$; however how to further improve the polarization fidelity remains an open question~\cite{gong2024isotope, gao2022nuclear, ru2023robust}.
Low temperature offers two potential advantages: $\sim 100$x longer \vbm electronic spin lifetime $T_1$ and stronger hyperfine interaction strength.

To enable resonant spin exchange between electronic and nuclear spins (known as electronic spin level anti-crossing, esLAC)~\cite{gong2024isotope,gao2022nuclear,mathur2022excited}, we apply an external magnetic field $B_z \approx 760~$G under which the energy levels corresponding to \vbm excited states $|m_s = 0\rangle$ and $|m_s = -1\rangle$ are nearly degenerate (see Supplemental Material).
We fit the measured hyperfine resonances between \vbm $|m_s=0\rangle\leftrightarrow|m_s=+1\rangle$ in the ODMR spectra as a function of the laser power to extract the nuclear spin polarization fidelity (Fig.~\ref{fig:fig4}a).
We find that the nuclear spin polarization at both $10$~K and $300~$K saturate around $30\%$ with similar laser powers ($\sim10~$mW) (Fig.~\ref{fig:fig4}b), suggesting that the \vbm spin lifetime $T_1$ and hyperfine interaction strength are not the limiting sources for nuclear polarization. 

\figFour

The spin polarization may be limited from the spin non-conserving terms in the hyperfine interaction (see Supplemental Material), which are temperature insensitive. 
In addition,  the observed nuclear spin polarization of \vbm persists down to $10~$K, in sharp contrast to the case of NV centers where the dynamic nuclear polarization process is  substantially suppressed at $T\lesssim50~$K, due to the lack of thermal averaging between NV excited states originating from different orbitals~\cite{fischer2013optical, block2021optically}. 
For \vbm, the lifting of the double degeneracy of excited states leads to a sufficiently large energy gap between the states, which renders the system insensitive to dynamic Jahn–Teller distortions~\cite{mathur2022excited}.

\emph{Conclusion and Outlook}---
In summary, our research offers a comprehensive investigation of the temperature-dependent spin-phonon interaction of \vbm centers in isotopically purified \hbnns.
Our results will be of critical importance in designing experiments to make use of \vbm centers as nanoscale thermometers~\cite{kucsko2013nanometre, choi2020probing, fujiwara2020real}.
While we focus on \vbmns, the methodologies developed here can be readily applied to a broad family of spin defects in two-dimensional materials \cite{stern2023quantum, guo2023coherent, li2022carbon, lee2022spin}

In addition, our results pave the way for a number of important directions beyond thermometry. 
First, owing to the 2D nature of the host hBN material, the local spin-phonon interaction of \vbm is expected to be strongly influenced by interfacial stacking. 
Therefore, \vbm centers in a few-layer hBN~\cite{durand2023optically} can potentially offer a brand-new quantum sensing modality --- probing interfacial phonon interactions in 2D devices.
Second, leveraging the 2D interfaces can provide the opportunity to engineer the local spin-phonon interactions, thereby further enhancing the spin properties of \vbmns.
We emphasize that these two new directions cannot be pursued by using NV centers in 3D diamond, as the spin-phonon interaction there is confined within the bulk diamond lattice and not expected to be influenced by the diamond surface.

\vspace{2mm}

\emph{Acknowledgements}:
We gratefully acknowledge Vincent Jacques, Tongcang Li, Shankar Mukherji, Erdong Song, Xi Wang and Norman Yao for helpful discussions.
This work is supported by NSF ExpandQISE 2328837 and the Center for Quantum Leaps of Washington University.
B. Huang, Y. Jin and G. Galli acknowledge the support of Air Force Office of Scientific Research (AFOSR) under Contract No. FA9550-22-1-0370, the Next Generation Quantum Science and Engineering (Q-NEXT) hub supported by the U.S. Department of Energy, Office of Science, National Quantum Information Science Research Centers and resources of the University of Chicago Research Computing Center.
Support for hBN crystal growth (E. Janzen and J. H. Edgar) was provided by the Office of Naval Research, award no. N00014-22-1-2582.

\vspace{1mm}

\bibliographystyle{apsrev4-1}
\bibliography{ref}

\end{document}


\title{Supplemental Material: \texorpdfstring{\\}{} Temperature Dependent Spin-phonon Coupling of Boron-vacancy Centers in Hexagonal Boron Nitride}

\author{
Zhongyuan~Liu,$^{1,*}$ 
Ruotian~Gong,$^{1,*}$ 
Benchen~Huang,$^{2,*}$ 
Yu~Jin,$^{2}$ 
Xinyi~Du,$^{1}$
Guanghui~He,$^{1}$\\
Eli~Janzen,$^{3}$
Li~Yang,$^{1}$
Erik Henriksen,$^{1}$
James Edgar,$^{3}$
Giulia Galli,$^{2,4,5}$
Chong~Zu$^{1,6,7,\dag}$
\\
\medskip
\normalsize{$^{1}$Department of Physics, Washington University, St.~Louis, MO 63130, USA}\\
\normalsize{$^{2}$Department of Chemistry, University of Chicago, IL 60637, USA}\\
\normalsize{$^{3}$Tim Taylor Department of Chemical Engineering, Kansas State University, Manhattan, KS, 66506, USA}\\
\normalsize{$^{4}$Pritzker School of Molecular Engineering, University of Chicago, IL 60637, USA}\\
\normalsize{$^{5}$Material Science Division, Argonne National Lab, IL 60439, USA}\\
\normalsize{$^{6}$Center for Quantum Leaps, Washington University, St. Louis, MO 63130, USA}\\
\normalsize{$^{7}$Institute of Materials Science and Engineering, Washington University, St. Louis, MO 63130, USA}\\
\normalsize{$^*$These authors contributed equally to this work}\\
\normalsize{$^\dag$To whom correspondence should be addressed; E-mail: zu@wustl.edu}\\
}

\date{\today}

\maketitle

\tableofcontents

\section{Experimental Setup}

We characterize the temperature-dependent spin properties of \vbm center in both \hbn and \hbnnat using a combination of a home-built confocal scanning microscope and a commercial close-cycle cryostat (Fournine Design) with optical access.
A 532~nm laser (Millennia eV High Power CW DPSS Laser) is used for \vbm spins initialization as well as detection.
The laser is shuttered by an acousto-optic modulator (AOM, G$\&$H AOMO 3110-120) in a double-pass configuration to achieve $>10^5:1$ on/off ratio.
An objective lens (Mitutoyo Plan Apo 20x/0.42 NA) focuses the laser beam to a diffraction-limited spot with diameter less than $1~\mu$m and collects the \vbm fluorescence.
The fluorescence is then separated from the laser beam by a dichroic mirror and filtered through a long-pass filter before being detected by a single photon counting module (Excelitas SPCM-AQRH-63-FC).
The signal is processed by a data acquisition device (National Instruments USB-6343).
The objective lens is mounted on a piezo objective scanner (Physik Instrumente PD72Z1x PIFOC) in order to control the position of the objective and scans the laser beam vertically.
The lateral scanning is performed by an X-Y galvanometer (Thorlabs GVS212) with a 4$f$ telescope.

The microwave driving field is generated from a signal generator (Stanford Research Systems SG384, SG386), amplified by a high-power amplifier (Mini-Circuits ZHL-15W-422-S+) and shuttered by a switch (Minicircuits ZASWA-2-50DRA+) to prevent any potential leakage.
All equipment are gated through a programmable multi-channel pulse generator (SpinCore PulseBlasterESR-PRO 500) with 2~ns temporal resolution.

The cryostat is cooled down to base temparature below 4~K by a closed-cycle helium compressors (Sumitomo F-40L).
A cryogenic temperature controller (LakeShore Model 336) and a sample temperature sensor with build-in heaters are used to maintain samples at desired temperature in range from 4~K to 350~K.

\begin{figure*}[b!]
 \centering
 \includegraphics[width=0.80\linewidth]{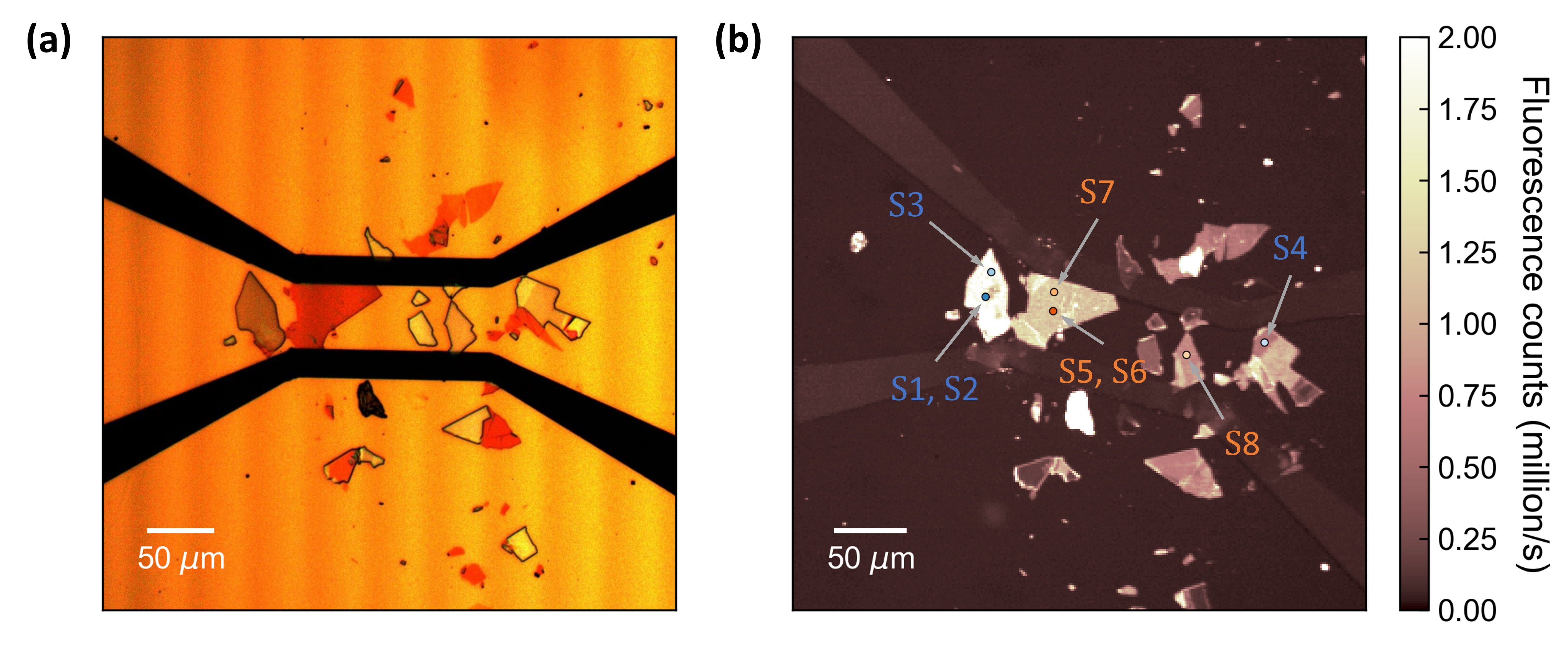}
 \caption{
 {\bf Optical and fluorescence images of the hBN flakes.}
 (a) A optical image of the hBN flakes on the coplanar waveguide.
 (b) A confocal scanning fluorescence image of the hBN flakes.
 The eight sets of data taken at six different spacial points are labeled as S1-S8, where S1-S4 are on \hbn flakes and S5-S8 are on \hbnnat flakes.
 }
 \label{fig:figS1}
\end{figure*}

\section{hBN Device Fabrication}

\subsection{\texorpdfstring{Growth of Isotopically Purified \hbnns}{}}
High-quality \hbn flakes are grown via precipitation from a molten Ni-Cr flux as detailed in \cite{janzen2023boron}. 
Ni, Cr, and isotopically pure \bten powders in the mass ratio 12:12:1 [Ni:Cr:B] are first loaded into an alumina crucible, then heated at 1550°C for 24 hours in a continuously flowing Ar/$\mathrm{H}_2$ atmosphere to melt all three components into a homogenous ingot and remove oxygen impurities. 
Next, this ingot is heated at 1550 °C for 24 hours in a static ${}^{15}\mathrm{N}_{2}$/$\mathrm{H}_2$ atmosphere to saturate the molten flux with ${}^{15}\mathrm{N}$, then slowly cooled at 1°C/hr to precipitate hBN from the solution. 
Since isotopically pure ($>99\%$) \bten and \nfive are the only sources of boron and nitrogen in the system, the hBN single crystals that precipitated have the same isotopes.

\subsection{\texorpdfstring{Creation of \vbm}~ defects}
We first exfoliate hBN nanosheets using tapes and then transfer them onto the Si/SiO$_2$ wafer. 
The wafer is pretreated with O$_2$ plasma at $50~$W for 1 minute (flow rate $20~$sccm).
The tapes and wafers are heated to $100^\circ$C for 1 minute before samples are carefully peeled off for maximum hBN flake yield \cite{huang2015reliable}. 
The wafers with \hbnnat and \hbn are then sent to CuttingEdge Ions LLC for $^4$He$^+$ ion implantation with energy $3~$keV and dose density $1~\mathrm{ion}/\mathrm{nm}^{2}$ to create \vbm defects.
After implantation, the hBN flakes with thickness ranging from $29~$nm to $78~$nm are transferred onto the coplanar waveguide using polycarbonate (PC) stamps.
Specifically, the temperature is raised in a stepwise manner to facilitate the transfer of hBN flakes. After successful transfer, the waveguide is immersed in chloroform for cleaning, removing any residual PC films.

\subsection{Fabrication of Coplanar Waveguide}
An impedance-matched ($50~\Omega$) coplanar waveguide is fabricated onto a $400~\mu$m thick sapphire wafer for microwave delivery (Supplementary Figure~\ref{fig:figS1}).
Specifically, photoresist layers (LOR 1A/S1805) are first applied to the sapphire wafer at a spin rate $4000~$rpm for 1 minute.
A carefully designed waveguide pattern with a $50~\mu$m wide central line is then developed using a direct-write lithography system (Heidelburg DWL66+), followed by an O$_2$ plasma cleaning process to remove resist residue. 
A $8~$nm chromium adhesion layer and a $180~$nm gold layer are deposited using thermal evaporation, followed by a lift-off process.

\section{Data Analysis}

\subsection{ODMR Spectrum Fitting}
As we discussed in the main text, the resonances of ODMR spectrum can be divided into two groups, corresponding to $|m_s = 0\rangle \leftrightarrow |m_s = -1\rangle$ and $|m_s = 0\rangle \leftrightarrow |m_s = +1\rangle$ transitions respectively.
Each group consists of four resonant peaks in \hbn samples and seven peaks in \hbnnat samples, due to the different nuclear spin numbers in the nitrogen isotopes.
Hence, each ODMR spectrum is modeled by two groups of equal-spaced Lorentzian profiles.
The fitting parameters of each group are transition frequency, Lorentzian width, and peak amplitude with ratio $1:3:3:1$ for \hbn sample (see section~\ref{quant_polarization}) or no constraint for \hbnnat sample.
An additional parameter, hyperfine splitting, is included to ascribe the interval between neighboring resonance in one group.
As a result, the ZFS, $D(T)$, is extracted from the average of two transition frequencies, while the hyperfine splitting, $A_{zz}(T)$, is directly obtained from the fitting.

\subsection{ZFS Temperature Dependence Fitting}
\begin{figure*}[h!]
 \centering
 \includegraphics[width=0.9\linewidth]{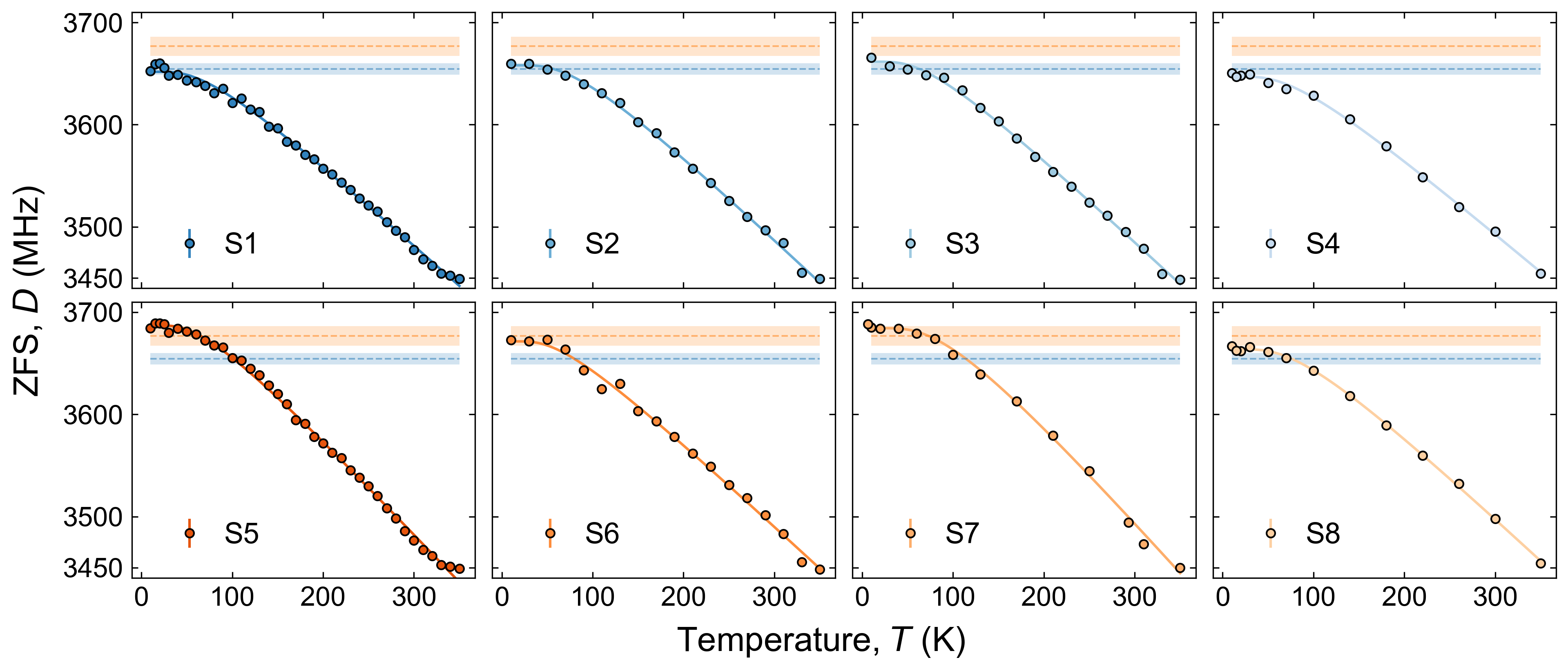}
 \caption{
 {\bf Extracted ZFS temperature dependence of \vbm at different spacial points in hBN flakes.}
 The average fitted value and corresponding uncertainty of one standard deviation of ZFS at $T=0$~K of \vbm in \hbn and \hbnnat flakes are represented as blue and orange shaded area respectively.
 }
 \label{fig:figS2}
\end{figure*}
We describe the measured ZFS of \vbm using the theoretical phonon model proposed in main text (Equation~3).
For both \hbn and \hbnnat samples, we take four sets of data at different positions on multiple flakes showing in Fig.~\ref{fig:figS1}, in order to reduce the experiment error.
Each data set is fitted independently with parameters $D_0$, $c_D$, and $\hbar\omega$, and summarized in Fig.~\ref{fig:figS2}.
The parameters are then be averaged and presented in Table~1 in the main text, with standard deviation across data sets.
We denote the averaged fitted value of ZFS at $T = 0$~K of \vbm in \hbn and \hbnnat flakes with blue and orange shaded area respectively, indicating the different ZFS behaviors of \vbm at very low temperature.

\subsection{Temperature Susceptibility}
We extract the temperature susceptibility of \vbm at room temperature from our experimentally measured ZFS around 300~K.
Specifically, in the high temperature region ($T > 200$~K), the ZFS of \vbm reveals a linear response to the temperature change.
We fit the ZFS with a linear function of the temperature in range from 250 to 350~K, and take the slope of that function as our determined susceptibility at room temperature.
The results for \vbm in both \hbnnat and \hbn are then averaged across four independent measurement at different spacial position on multiple flakes showing in Fig.~\ref{fig:figS1}, while the uncertainty is determined by the standard deviation.

\subsection{Thermometry Sensitivity}
Due to its susceptibility to external environment like magnetic field as well as the electric field, temperature, and pressure, the ZFS of \vbm can be an important addition to the tool set of quantum sensing \cite{gottscholl2021spin}.
The Zeeman splitting between $|m_s = -1\rangle$ and $|m_s = +1\rangle$ can be directly probed via ODMR, and the corresponding static magnetic field sensitivity takes the form \cite{dreau2011avoiding,barry2020sensitivity}
\begin{equation} \label{eqDCSensitivity}
\begin{split}
\eta_\mathrm{B} \approx\frac{2 \pi}{\gamma_e \sqrt{R}} (\max|\frac{\partial{C(\nu)}}{\partial{\nu}}|)^{-1} \approx \frac{8\pi}{3\sqrt{3}} \frac{1}{\gamma_e}\frac{\Delta \nu} {C_m\sqrt{R}},
\end{split}
\end{equation}
where $\gamma_e$ denotes the electron gyromagnetic ratio ($2\pi\times 2.8~$MHz/G), $R$ the photon detection rate, $C(\nu)$ the ODMR measurement contrast at microwave frequency $\nu$, $C_m$ the maximum contrast, and $\Delta \nu$ the FWHM linewidth assuming a single Lorenztian resonance.
Here, directly comparing the maximum $\frac{\partial{C(\nu)}}{\partial{\nu}}$ of each spectrum more accurately reflects the sensitivity as the hyperfine levels largely overlap with each other in \hbnnat and one cannot resolve individual Lorenztian \cite{gong2024isotope}.
Although the underlying coupling mechanism is different, the temperature dependent ZFS can also be determined using the same ODMR technique.
Specifically, one needs to substitute $\gamma_e$ with the temperature susceptibility $\chi$ and omit corresponding $2\pi$ to account for the non-linear relationship resulting from spin-phonon relationship.
\begin{equation} \label{eqTSensitivity}
\begin{split} \eta_\mathrm{T} \approx \frac{1}{|\chi| \sqrt{R}} (\max|\frac{\partial{C(\nu)}}{\partial{\nu}}|)^{-1},
\end{split}
\end{equation}

\begin{figure}[h!]
    \centering
    \includegraphics[width=0.6\textwidth]{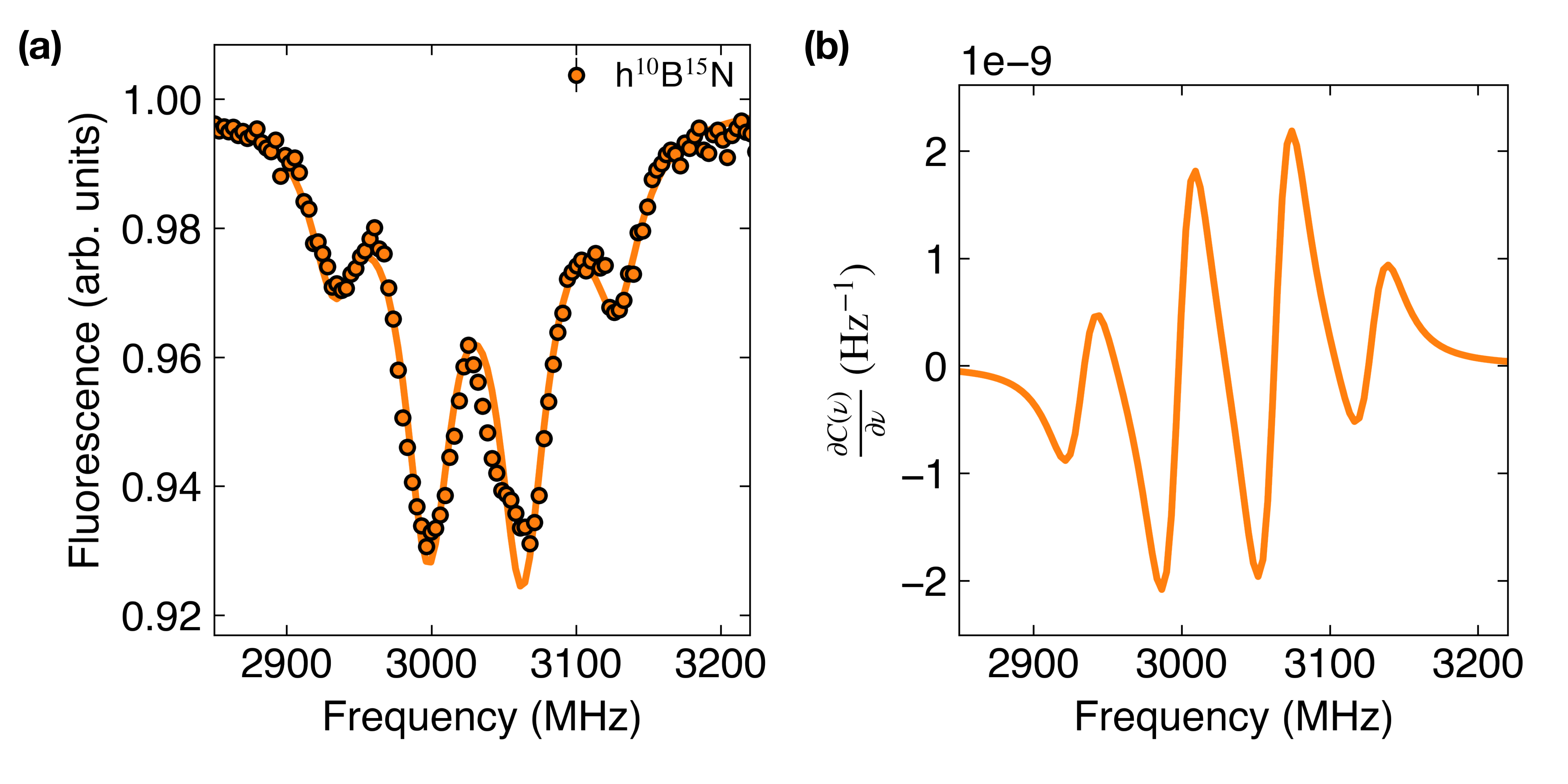}
    \caption{{\bf Optimization of temperature sensitivity for ODMR measurements.}
    (a) ODMR spectrum of \vbm in \hbn after optimizing the laser and microwave powers for better sensitivity.
    (b) Derivatives of the fitted ODMR spectrum in (a).}
    \label{Supp_sensitivity}
\end{figure}

Moreover, we remark that when performing the ODMR measurement in main text Figure~2a, we utilize small laser and microwave powers to avoid spectral broadening to resolve the intrinsic linewidth of the resonances. 
However, to optimize sensitivity, we may want to further fine-tune the laser and microwave powers.
Specifically, increasing laser power leads to a higher count rate $R$ but lower maximum contrast and broadened linewidth, while increasing microwave power can boost contrast but also cause power broadening of the resonances.
By carefully adjusting the laser ($\sim 0.5$ mW) and microwave power while monitoring the ODMR signal from a flake with thickness $\sim 64$ nm, we obtain the an improved spectrum for sensing (Fig~\ref{Supp_sensitivity}b).
After extracting the maximum $\frac{\partial{C(\nu)}}{\partial{\nu}}$ from now improved spectrum, we plug the obtained values ($\chi = -0.77\pm0.03$~MHz/K, $R = 2.6\times 10^6~$Hz, and $\max(|\frac{\partial{C(\nu)}}{\partial{\nu}}|) = 2.2 \times 10^{-9} ~\mathrm{Hz}^{-1}$) into Equation.~\ref{eqTSensitivity} and obtain $\eta_T \approx 0.37~ \mathrm{K}/\sqrt{\mathrm{Hz}}$ at room temperature.

Our estimation shows that the temperature sensitivity of \vbm centers in hexagonal boron nitride (hBN) is nearly an order of magnitude better than that achieved using a similar protocol on NV centers in nanodiamonds (50-100 nm in diameter), which typically range from $1.4$ to $2.3~\mathrm{K}/\sqrt{\mathrm{Hz}}$ \cite{fujiwara2020real, tzeng2015time,choi2020probing,fujiwara2021diamond, gu2023simultaneous}.
NV centers in nanodiamonds have been widely studied for biosensing due to their chemical inertness and biocompatibility. 
Over the past decade, extensive research on NV centers in nanodimaonds has led to advanced techniques like surface functionalization and quantum controls to optimize their performance.
While high sensitivity is a valuable feature, it may not translate directly into practical applications due to various experimental factors that can introduce artifacts~\cite{bradac2020optical}. However, the spin states of \vbm centers and NV centers are much more robust and controllable compared to metrics like fluorescence intensity or lifetime, which are more susceptible to interference.
Moreover, the promising initial performance of \vbm centers in hBN indicates significant potential for improvement, especially given the increasing research interest in this area.

To further illustrate the potential advantages of \vbm centers, it is important to also consider other characteristics of thermometers such as their thermal conductivity and how they will physical contact target objects.
Specifically, prior works exploring NV thermometry primarily make use of nanodiamonds instead of bulk diamond, where the heat dissipation process is constrained within small-scale.
In our study, the hBN flakes hosting the \vbm centers have a thickness ranging from $30-80$ nm, which is about four orders of magnitude smaller than typical bulk diamond samples ($\sim 300-500~\mu$m).

Moreover, hBN has a thermal conductivity, $\sim 500~$W/(m$\cdot$K)~\cite{cai2019high, yuan2019modulating}, much lower than that of diamond ($\sim 2500$~W/(m$\cdot$K))~\cite{graebner1995thermal}.
As a result, by attaching the hBN sensing sheet to the target sample, we expect the \vbm sensors at different locations to faithfully capture the local temperature distribution.
Moreover, unlike nanodiamond particles that can freely rotate, the \vbm spins in a thin sheet of hBN are all aligned along a predetermined direction (the out-of-plane direction), eliminating the need to calibrate the sensor orientation.
Therefore, we believe \vbm centers in hBN have the potential to outperform NV centers as nanoscale thermometers, offering a robust and highly sensitive tool for temperature measurement.

\section{AB INITIO CALCULATION}

\subsection{Electronic Structure}
The electronic structure calculations employ the projector-augmented-wave (PAW) method implemented in the open-source plane-wave-based Quantum ESPRESSO software package (QE)~\cite{giannozzi2009quantum, giannozzi2017advanced} with a kinetic energy cutoff of 75 Ry. Spin-unrestricted calculations are performed using the Perdew-Burke-Ernzerhof (PBE) functional~\cite{perdew1996generalized} in the computation of relaxed atomic geometries, phonons from the frozen-phonon approach~\cite{chaput2011phonon}, and hyperfine tensor $A$. The threshold of energy convergence is set to $10^{-8}$ eV, and that for force on atoms is set to 0.005 eV/\r{A} for geometry optimization calculations.

As discussed in the main text, the temperature dependence can be captured via phonons. We note that for Eq. (2) in the main text, only diagonal terms are considered for the second-order contributions, leaving out the off-diagonal terms. This simplification proves to be good for the study of NV$^-$ center in diamond~\cite{tang2023first} and $V^-_{\mathrm{B}}$ in hBN.

The phonon calculation for $V^-_{\mathrm{B}}$ in hBN is performed using a 287-atom supercell with only $\Gamma$-point sampling. The lattice parameters are taken from experiments at 0 K~\cite{paszkowicz2002lattice}. A total of 861 phonon modes are obtained, including 3 trivial modes corresponding to translations with no contribution to the temperature dependence. Therefore, all the 858 non-trivial phonon modes are then used to calculate the second-order derivative according to Eq. (2) in the main text, with the step of displacement $\delta q_i$ set to 0.1 \r{A}.

\begin{figure}[hbt!]
    \centering
    \includegraphics[width=0.9\textwidth]{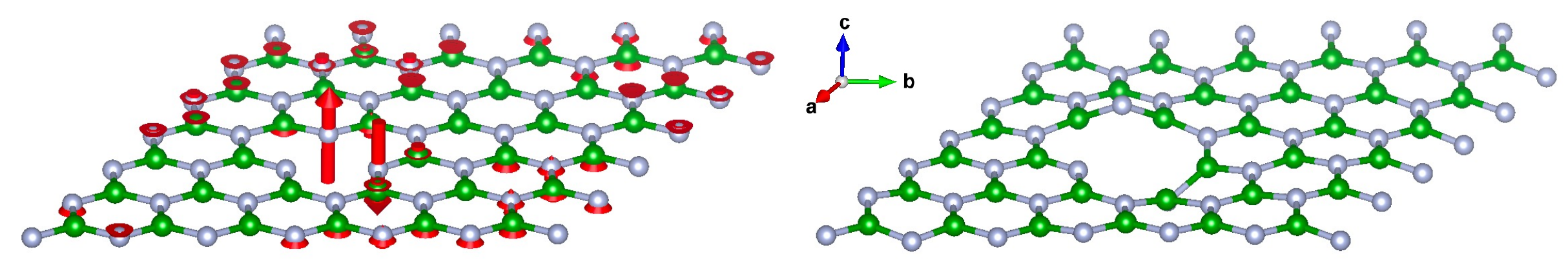}
    \caption{Representative vibrational mode with $\omega = 16$ meV identified from the temperature dependence of the hyperfine interaction. The red arrows on the left panel show the out-of-plane vibrational mode while the right panel shows the actual movement of the atoms. Note that only the sheet containing the vacancy is shown to better illustrate the displacement of atoms.}
    \label{fig:vibrational_mode}
\end{figure}

\subsection{Hyperfine Interaction}
The hyperfine interaction between nuclear spin and electron spin includes the isotropic (Fermi contact) term at the nucleus site $N$ and the anisotropic dipole-dipole interaction term near the nucleus site $N$
\begin{align}
    A_{\text{iso}}(N) & = \frac{2\mu_0}{3} g_e \mu_e g_N \mu_N \rho_{\text{spin}}(\textbf{R}),\\
    A_{\text{aniso}}(N) & = \frac{\mu_0}{4\pi} g_e \mu_e g_N \mu_N \int d^3 r \rho_{\text{spin}}(\textbf{r}) \frac{3\cos^2\theta - 1}{2r^3},
\end{align}
where $\mu_0$ is the permeability of vacuum, $g_e, g_N$ is the electron and nuclear Land\'e g-factor, $\mu_e, \mu_N$ is the Bohr and nuclear magneton. $\mathbf{r}$ is the displacement between the electron and the nucleus at $\textbf{R}$. $\rho_{\text{spin}}$ denotes the electronic spin density computed from DFT, and $\theta$ is the angle between $\mathbf{r}$ and $z$ axis.

\begin{figure}[hbt!]
    \centering
    \includegraphics[width=0.9\textwidth]{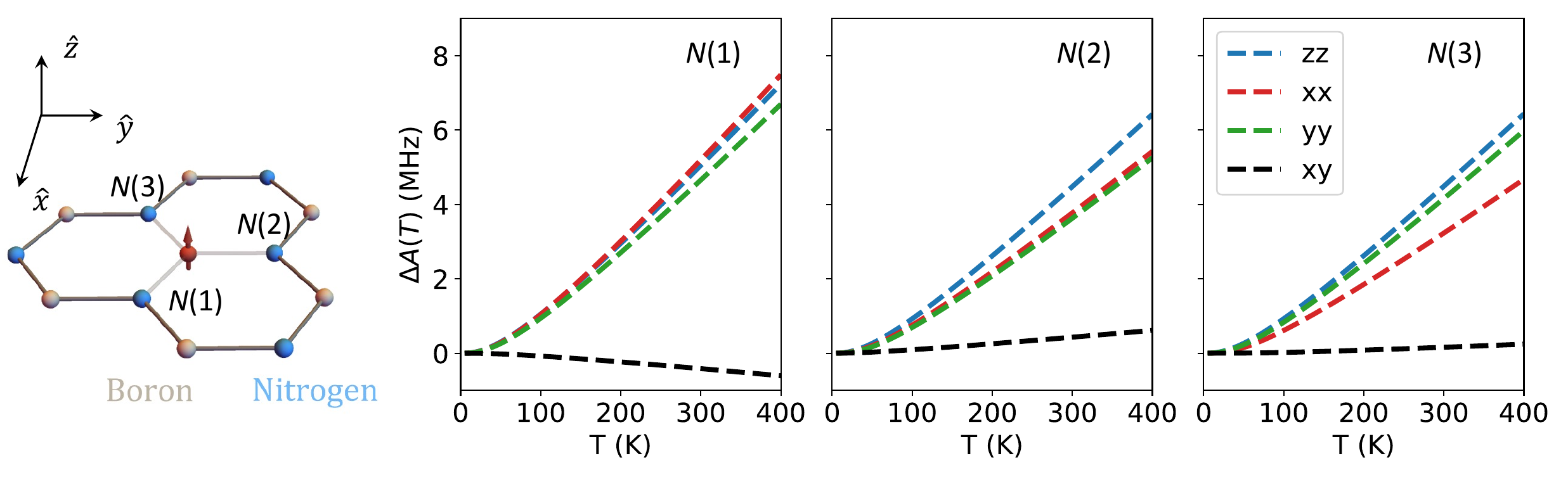}
    \caption{Temperature dependence of the full hyperfine tensor of three nearest \nfive computed from the second-order spin-phonon coupling.}
    \label{fig:A_temp_SI}
\end{figure}

The temperature dependence of the full hyperfine tensor of the three nearest \nfive atoms computed via the method discussed in the previous subsection are plotted in Fig.~\ref{fig:A_temp_SI}. Only second-order contribution is considered since the lattice expansion effect is negligible. Note that $A_{\mathrm{yz}} = A_{\mathrm{zx}} = 0$ due to the point group symmetry of the defect center.

\subsection{Zero Field Splitting}

The ZFS originates from the dipolar spin-spin interaction between electrons and can be computed as
\begin{equation}
    D_{ab} = \frac{\mu_0 g_e^2 \mu_e^2}{4\pi}\sum_{i<j}^{\text{occ.}} \chi_{ij} \bigg\langle \Psi_{ij}\bigg|\frac{r^2 \delta_{ab} - 3 r_a r_b}{r^5} \bigg|\Psi_{ij}\bigg\rangle,
\end{equation}
where $|\Psi_{ij}\rangle$ represents a two-particle Slater determinant constructed from the Kohn-Sham ground state and $\chi_{ij} = \pm 1$ when $i, j$ have the same/different spins. And the summation runs over all the possible electron pairs.

It has been noted by Iv\'ady et al.~\cite{ivady2020ab} that it's necessary to apply hybrid functionals~\cite{heyd2003hybrid} to a 971-atom supercell, together with a post-correction for spin contamination~\cite{biktagirov2020spin}, to obtain a reasonable estimation of ZFS. This calculation is computationally expensive and impractical for the investigation of temperature dependencies, which involves hundreds of single point calculations.

\begin{figure}[b!]
    \centering
    \includegraphics[width=0.7\textwidth]{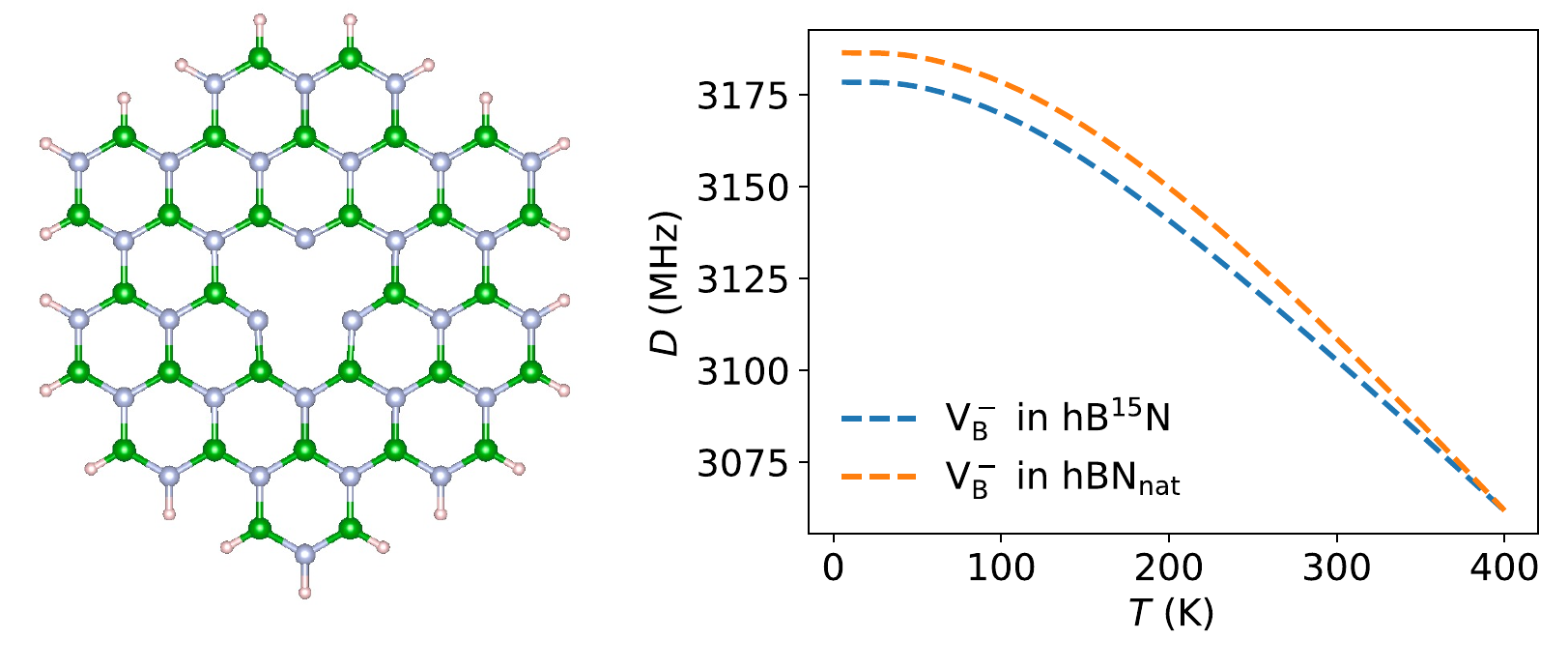}
    \caption{Temperature dependence of the zero field splitting of both hB$^{15}$N and \hbnnat (right), computed from the cluster model (left) by only considering the second-order coupling between spin and local phonon modes near the defect. Here, we aligned the two curves at around $400$ K for a better comparison with experiments.}
    \label{fig:D_temp_SI}
\end{figure}

To qualitatively understand this temperature dependence, we focus on the coupling between spin and local phonon modes around the \vbm center. Specifically, we cut a cluster (passivated with hydrogen atoms) with $7.5$ \AA\ radius~\cite{mondal2023spin} from the supercell, and apply local phonon displacements on top of it. Then we performed DFT calculations with the PBE0~\cite{adamo1999toward} functional and the def2-SVP basis set to compute the ground state and the ZFS using the ORCA package~\cite{neese2020orca}, and the results are shown in Fig.~\ref{fig:D_temp_SI}. It can be noticed that the absolute values of the computed ZFS and the magnitude of the variation as a function of temperature are underestimated compared to the experiments. This discrepancy could be attributed to i). ignoring the contributions of a large number of non-local phonon modes, and ii). approximating the supercell using the cluster model in the calculation of the ZFS. Despite these deficiencies, the decreasing trend in $D$ with increasing temperature is correctly captured by our present calculations, and the difference between the two isotopes is also well reproduced, which again solidifies the phenomenological model that we've been fitting throughout this work.

We also note that P\'eter et al.~\cite{udvarhelyi2023planar} have studied the effects of lattice strain on the ZFS. From Supplementary Fig. 2 of Ref.~\cite{udvarhelyi2023planar}, we can extract that the first-order contribution to the ZFS's temperature dependence is also negligible, which aligns with our expectations. Their supplementary Fig. 4b also corroborates our conclusion that the first-order contribution to the hyperfine tensor is negligible.

\subsection{Spin Relaxation Time}

The spin relaxation of a defect could be credited to multiple possible mechanisms. Different mechanisms could play important roles in different temperature ranges. Since the ZFS of $V^-_{\mathrm{B}}$ in hBN is $\sim3.5$ GHz, it's well below the typical phonon energies of hBN~\cite{mondal2023spin}. For temperatures well above 0 K, the one-phonon (direct) process can be safely ignored. Therefore, we mainly focus on the two-phonon processes. For the standard Orbach process~\cite{orbach1961spin}, it's also unlikely to happen for $V^-_{\mathrm{B}}$, because the lowest-lying excited state is $\sim800$ meV higher than the triplet ground state~\cite{reimers2020photoluminescence}, well beyond the phonon cutoff frequency which is $\sim200$ meV in hBN~\cite{reich2005resonant}.
 
One other possible two-phonon process is the Raman scattering process and it could be driven by either first-order or second-order spin-phonon interactions. Ref.~\cite{cambria2023temperature} proposed a criterion to distinguish which of the two is dominant, namely the ratio between these two contributions, i.e., $(2\pi D/ \omega)^2$, where $D$ is the ZFS and $\omega$ is the acoustic phonon energy. Ref.~\cite{cambria2023temperature} also showed that for NV$^-$ center in diamond, $(2\pi D/ \omega)^2 \sim 10^{-7}$ and therefore the dominant driving force is second-order spin-phonon interactions. For $V^-_{\mathrm{B}}$ in hBN, we can also estimate its value as $10^{-5}$, where we've taken the acoustic phonon energy as $\sim26$ meV~\cite{krummheuer2002theory}. We can, therefore, conclude safely that the major mechanism responsible for the spin relaxation in $V^-_{\mathrm{B}}$ is the Raman scattering driven by second-order spin-phonon interactions, and our analysis is corroborated by Ref.~\cite{mondal2023spin}.

In this scenario, the spin relaxation rate can be modeled as
\begin{equation}
    \Gamma(T) = 1/T_1 = \sum_{i} A_i n_i (n_i + 1) + A_S,
\end{equation}
which is a more general form than Eq. 4 in the main text. Since we only identified a single representative phonon mode for the temperature dependence of ZFS, we only fit a single set of $A, n$.

\begin{figure}[h]
 \centering
 \includegraphics[width=0.5\textwidth]{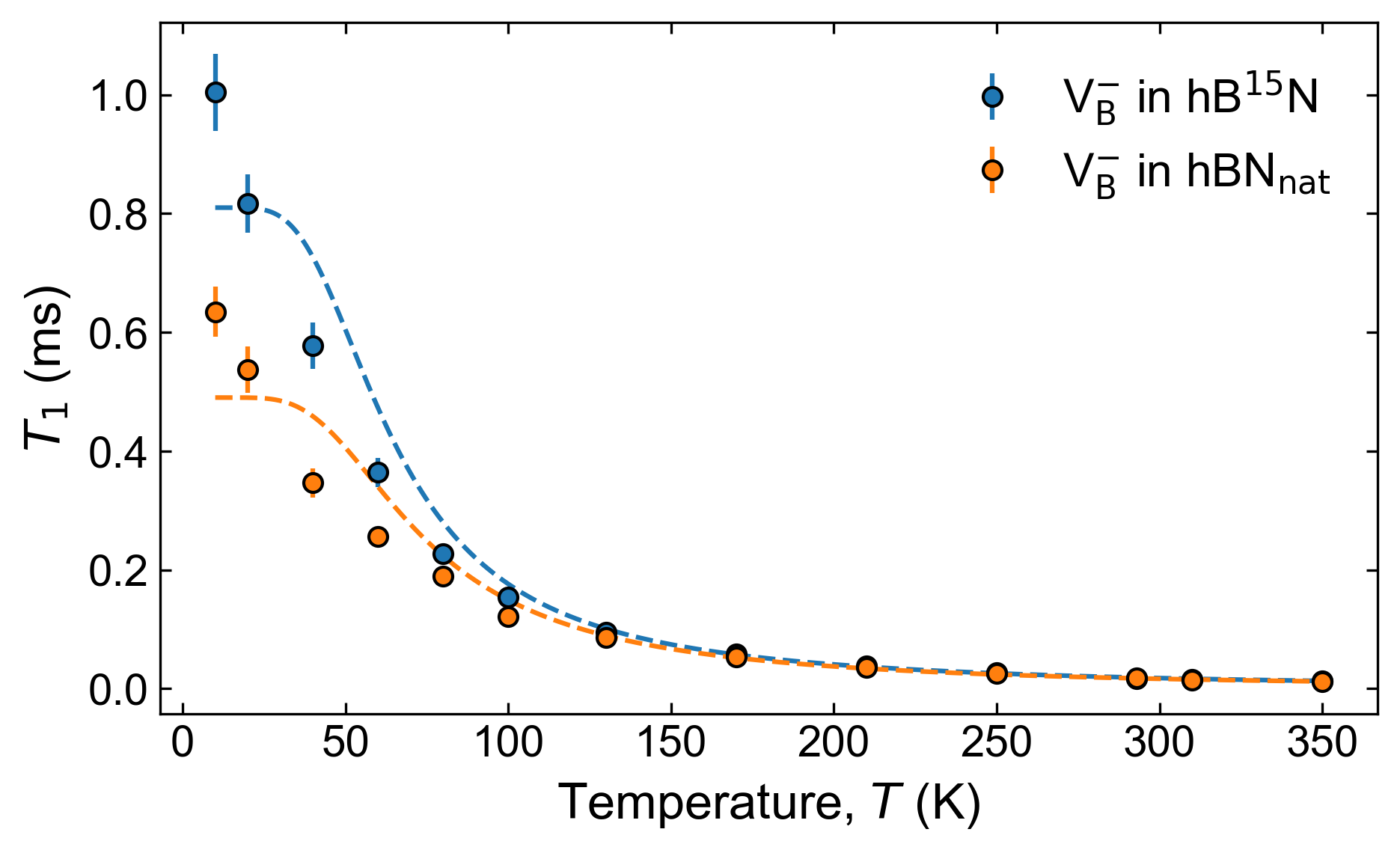}
    \caption{{\bf Spin relaxation time $T_1$ of \vbm as a function of temperature in range 10-350~K.}
    The dotted lines represent the $T_1$ temperature-dependence fitted by the model assuming two-phonon Raman process (Eq.~4 in main text).
    }
 \label{fig:figR1}
\end{figure}

\section{ODMR Contrast and Microwave Loss}

In main text we mention the ODMR contrast exhibits a maximum around 210~K for \vbm in both hBN samples, and here we will demonstrate that this contrast feature is not correlated with the microwave loss during transmission.
In a typical ODMR measurement, the transition amplitude (contrast) is depend on both laser intensity and microwave strength.
While with fixed laser intensity, the ODMR contrast will decrease with lower microwave power \cite{dreau2011avoiding}.
\begin{figure*}[h!]
 \centering
 \includegraphics[width=0.95\linewidth]{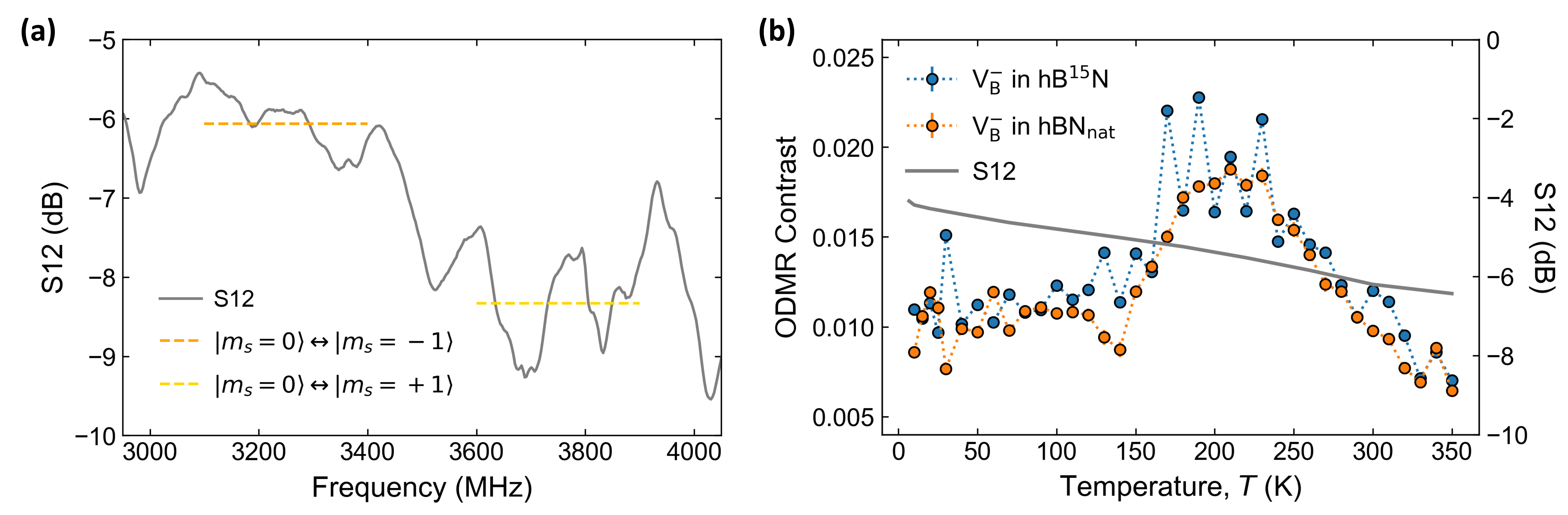}
 \caption{
 {\bf S-parameter transmission efficiency test (S12) results.}
 (a) Coplanar waveguide transmission efficiency frequency response at room temperature.
 The dotted lines represent the average transmission efficiency in frequency range 3100-3400~MHz ($|m_s=0\rangle \leftrightarrow |m_s=-1\rangle$) and 3600-3900~MHz ($|m_s=0\rangle \leftrightarrow |m_s=+1\rangle$) respectively.
 The larger microwave loss in $|m_s=0\rangle \leftrightarrow |m_s=+1\rangle$ transition leads to it's smaller contrast compared to $|m_s=0\rangle \leftrightarrow |m_s=-1\rangle$ transition.
 (b) ODMR contrast and S12 test with temperature dependence.
 The ODMR contrasts of the $|m_s=0\rangle \leftrightarrow |m_s=-1\rangle$ transition of \vbm in \hbn and \hbnnat flakes are labeled with blue and orange respectively.
 The contrast reaches its maximum around 210~K for both isotopes.
 The S12 value is taken by the average result in 3100-3500~MHz frequency range which covers the resonant frequency of the $|m_s=0\rangle \leftrightarrow |m_s=-1\rangle$ transition from 10~K to 350~K (Fig.~1c,d in main text).
 }
 \label{fig:figS6}
\end{figure*}

We perform an S-parameter transmission efficiency test (S12) on our coplanar waveguide in the cryostat using a vector network analyzer (Agilent Technologies N5239A) within temperature range from 10~K to 350~K.
The transmitted power is reduced by 4-6.5~dB in the frequency range from 3100~MHz to 3500~MHz which covers the resonant frequency of the $|m_s=0\rangle \leftrightarrow |m_s=-1\rangle$ transition in our experiment (Fig.~1c,d in main text).
We take the average value in the frequency range and plot it as a gray line in Fig.~\ref{fig:figS6}(b).
The transmission efficiency result exhibits a monotonic decrease with increasing temperature.
In comparison, the ODMR contrasts of the $|m_s=0\rangle \leftrightarrow |m_s=-1\rangle$ transition of \vbm in \hbn and \hbnnat flakes firstly increases, peaks around 210~K and then decreases as temperature going up.
As a result, the ODMR contrast behavior of \vbm can not be explained by the transmission loss in this case.
We note that further investigation is required to understand the mechanism of the \vbm ODMR contrast temperature dependence.

\section{Nuclear Spin Polarization}

\subsection{\texorpdfstring{Interaction Hamiltonian for $\mathrm{V}_\mathrm{B}^-$}{} and nuclear spins}
To understand the nuclear spin polarization process, we must first derive the Hamiltonian that governs the \vbm and nuclear spins interaction. 
The \vbm electronic spin-1 operators can be written as
\begin{equation} \label{vbop}
\begin{split}
S_z =
\begin{bmatrix}
1 & 0 & 0\\
0 & 0 & 0\\
0 & 0 & -1
\end{bmatrix}	,~
S_x = \frac{1}{\sqrt{2}}
\begin{bmatrix}
0 & 1 & 0\\
1 & 0 & 1\\
0 & 1 & 0
\end{bmatrix}	,~
S_y = \frac{1}{\sqrt{2}i}
\begin{bmatrix}
0 & 1 & 0\\
-1 & 0 & 1\\
0 & -1 & 0
\end{bmatrix}   .
\end{split}
\end{equation}
Here we define the spin raising and lowering operators for the \vbm electron spin as
\begin{equation} \label{vbladder}
S_+ = \sqrt{2}
\begin{bmatrix}
0 & 1 & 0\\
0 & 0 & 1\\
0 & 0 & 0
\end{bmatrix}
= S_x + iS_y ,~
S_- = \sqrt{2}
\begin{bmatrix} 
0 & 0 & 0\\
1 & 0 & 0\\
0 & 1 & 0
\end{bmatrix}
= S_x - iS_y,
\end{equation}
and consequently the \vbm spin operators can be expressed as
\begin{equation} \label{vbopladder}
S_x = \frac{S_+ + S_-}{2} ,~
S_y = \frac{S_+ - S_-}{2i} .
\end{equation}

Similarly, for \nfive nuclear spins, we have operators
\begin{equation} \label{nucop}
\begin{split}
I_z = \frac{1}{2}
\begin{bmatrix}
1 & 0 \\
0 & -1
\end{bmatrix}	,~
I_x = \frac{1}{2}
\begin{bmatrix}
0 & 1 \\
1 & 0
\end{bmatrix}	,~
I_y = \frac{1}{2i}
\begin{bmatrix}
0 & 1 \\
-1 & 0
\end{bmatrix}   .
\end{split}
\end{equation}
\begin{equation} \label{nucladder}
I_+ =
\begin{bmatrix}
0 & 1 \\
0 & 0
\end{bmatrix}
= I_x + iI_y ,~
I_- =
\begin{bmatrix}
0 & 0 \\
1 & 0
\end{bmatrix}
= I_x - iI_y ,
\end{equation}
\begin{equation} \label{nucopladder}
I_x = \frac{I_+ + I_-}{2} ,~
I_y = \frac{I_+ - I_-}{2i} .
\end{equation}

The hyperfine interaction term between \vbm and three \nfive nuclear spins takes the form $\sum_{j=1}^{3}\mathbf{S}\mathbf{A}^{j}\mathbf{I}^j$
where the hyperfine parameters tensor $\mathbf{A}$ for a single nuclear spin can be expressed as
\begin{equation} \label{hyptensor}
\mathbf{A} =
\begin{bmatrix}
A_{xx} & A_{xy} & A_{xz} \\
A_{yx} & A_{yy} & A_{yz} \\
A_{zx} & A_{zy} & A_{zz}
\end{bmatrix}
=
\begin{bmatrix}
A_{xx} & A_{xy} & 0 \\
A_{yx} & A_{yy} & 0 \\
0      & 0      & A_{zz}
\end{bmatrix}.
\end{equation}
Here the $\hat{z}$-axis is defined along the c-axis of hBN (perpendicular to the lattice plane, see main text Figure.~1), $\hat{x}$ and $\hat{y}$ lie in the lattice plane, with $\hat{x}$ oriented along one of the three in-plane nitrogen bonds. Due to the mirror symmetry of \vbm with respect to the $\hat{x}-\hat{y}$ plane, the four terms $A_{xz}$ $ A_{yz}$ $ A_{zx}$ $ A_{zy}$ vanish \cite{ivady2020ab, gao2022nuclear}. We can then expand the hyperfine interacting Hamiltonian to its full form
\begin{equation} \label{hypderive}
\begin{split}
    \sum_{j=1}^{3}\mathbf{S}\mathbf{A}^{j}\mathbf{I}^j & = \sum_{j=1}^{3} (A_{zz}^{j}S_z I_z^j + A_{xx}^{j}S_x I_x^j +A_{yy}^{j}S_y I_y^j +A_{xy}^{j}S_xI_y^j +A_{yx}^{j}S_yI_x^j ) \\
    & = \sum_{j=1}^{3} [A_{zz}^{j}S_z I_z^j + \frac{1}{4}A_{xx}^{j}(S_+ + S_-)(I_+^j+I_-^j)-\frac{1}{4i}A_{yy}^{j}(S_+ - S_-) (I_+^j-I_-^j) \\
    & ~~~ + \frac{1}{4i}A_{xy}^{j}(S_+ + S_-) (I_+^j-I_-^j) + \frac{1}{4i}A_{yx}^{j}(S_+ - S_-) (I_+^j+I_-^j)] \\ 
    & = \sum_{j=1}^{3} [A_{zz}^{j} S_z I_z^j + \dfrac{A_{xx}+A_{yy}}{4} (S_+I_-^j  + S_-I_+^j) \\
    & ~~~ +  (\dfrac{A_{xx}-A_{yy}}{4} + \dfrac{A_{xy}}{2i}) S_+I_+^j + (\dfrac{A_{xx}-A_{yy}}{4} - \dfrac{A_{xy}}{2i}) S_-I_-^j].
\end{split}
\end{equation}
Note that $A_{xy} = A_{yx}$ from the symmetry in the last step\cite{ivady2020ab}. Finally, since $(S_+I_-^j)^\dag = (S_-I_+^j)$ and $(S_-I_-^j)^\dag = (S_-I_-^j)$, we can further simplify the expression to the form
\begin{equation} \label{hyp}
    \sum_{j=1}^{3}\mathbf{S}\mathbf{A}^{j}\mathbf{I}^j = \sum_{j=1}^{3} [A_{zz}^{j}S_z I_z^j + (A_1^{j}S_+ I_-^j + h.c.) + (A_2^{j}S_+ I_+^j + h.c.)],
\end{equation}
where we define $A_1^j = \frac{1}{4}(A_{xx}^j+A_{yy}^j)$ and $A_2^j = \frac{1}{4} (A_{xx}^j-A_{yy}^j)+\frac{1}{2i}A_{xy}^j$.

\subsection{Spin population transfer near esLAC}

To enable resonant spin exchange between electronic and nuclear spins, we apply an external magnetic field $B_z \approx 760~$ under which the energy levels corresponding to \vbm excited state $|m_s = 0\rangle$ and $|m_s = -1\rangle$ are nearly degenerate (known as electronic spin level anti-crossing, esLAC)~\cite{gong2024isotope,gao2022nuclear,mathur2022excited}. The hyperfine interaction Hamiltonian is in full display as written in the previous section since the secular approximation is no longer valid.
In the equation \ref{hyp}, the more dominant term $(A_1^{j}S_+ I_-^j + h.c.)$ leads to electron-nuclear spin flip-flop, $|m_s=0, m_I=
\:\downarrow \rangle \leftrightarrow |m_s=-1, m_I=\:\uparrow\rangle$, while the secondary term $(A_2^{j}S_+ I_+^j + h.c.)$ connects the other two states, $|m_s=0, m_I=\:\uparrow \rangle \leftrightarrow |m_s=-1, m_I=\:\downarrow \rangle$ (main text Fig.~4b inset).

As a result, the \nfive nuclear spin population will preferentially transfer to $|m_I = \:\uparrow\rangle$ state when a strong optical polarization continuously pumps the electronic spin from $|m_s=\pm1\rangle$ to $|m_s=0\rangle$.

\subsection{Quantifying polarization} \label{quant_polarization}
To quantitatively describe the nuclear spin polarization as a function of temperature, we record the ODMR spectra while sweeping the laser power, i.e., the optical pumping rate.
We extract the nuclear spin polarization of the nearest three \nfive by first assuming each nuclear spin is individually prepared to $|m_I=\:\uparrow \rangle$ with probability $P(\ket{\uparrow})$ and $|m_I=\:\downarrow \rangle$ with probability $1-P(\ket{\uparrow})$.
The energy sublevels accounting for the three \nfive spins should effectively follow a bionomial distribution with $n=3$, the total number of a sequence of independent events.
For a random variable $X$ defined as the number of \nfive nuclear spins being measured to be $|m_I=\:\uparrow \rangle$ in a single observation, the probability mass function is
\begin{equation} \label{eq5}
    \mathrm{Pr} (X = k) = \binom{n}{k}P^k(1-P)^{3-k}.
\end{equation}
For example, the four possible nuclear spin configurations correspond to $\sum m_I = {-\frac{3}{2}, -\frac{1}{2}, \frac{1}{2}, \frac{3}{2}}$ and $X = {0, 1, 2, 3}$.
Therefore, the theoretical probability distribution follows $\{\binom{3}{3} (1-P)^3 : \binom{3}{2} P(1-P)^2 : \binom{3}{1} P^2(1-P) : \binom{3}{0} P^3 \} = \{(1-P)^3 :3 P(1-P)^2 : 3 P^2(1-P) : P^3\}$. 
When there is no nuclear spin polarization ($P=0.5$), we obtain amplitudes $\{1:3:3:1\}$ for the ODMR spectra.

To quantitatively extract the nuclear polarization, we first fit the ODMR spectrum using the sum of four equally spaced Lorentzian distribution whose amplitude follows the aforementioned ratio.
However, we note that the gyromagnetic ratio for \nfive is negative, which reverses the order of the sublevels in the ODMR spectrum. 
In this case, for the \vbm transition between electronic $|m_s=0\rangle$ and $|m_s = +1\rangle$ states, the lowest-frequency resonance corresponds to $\sum m_I = \frac{3}{2}$ or $X=3$.

\bibliographystyle{apsrev4-1}
\bibliography{ref.bib}